\documentclass[a4paper,fleqn]{cas-dc}

\usepackage[authoryear,longnamesfirst]{natbib}

\usepackage{hyperref}
\usepackage{booktabs}
\usepackage{color,array}
\usepackage{graphicx}
\usepackage{multirow}

\usepackage{url}
\usepackage{algorithmic}
\usepackage{algorithm}
\usepackage{makecell}
\usepackage{amsthm}
\usepackage{amssymb}
\usepackage{multirow}
\usepackage[caption=false,font=footnotesize]{subfig}
\usepackage{amsfonts} 

\def\tsc#1{\csdef{#1}{\textsc{\lowercase{#1}}\xspace}}
\tsc{WGM}
\tsc{QE}
\begin{document}
\let\WriteBookmarks\relax
\def\floatpagepagefraction{1}
\def\textpagefraction{.001}

%\shorttitle{<short title of the paper for running head>} 
\shorttitle{Network Collaborator}

%\shortauthors{<short author list for running head>}  
\shortauthors{Kai Wu et al.}  
\title [mode = title]{Network Collaborator: Knowledge Transfer Between Network Reconstruction and Community Detection}  

%\tnotemark[<tnote number>] 

%\tnotetext[<tnote number>]{<tnote text>} 


\author[1]{Kai Wu}[type=editor]
%\author[<aff no>]{<author name>}[<options>]

%\cormark[1]

%\fnmark[<footnote mark no>]

% \ead{kwu@xidian.edu.cn}

%\ead[url]{<URL>}

%\credit{<Credit authorship details>}
%, Xi'an 710071, China
\affiliation[1]{organization={School of Artificial Intelligence, Xidian University},
            %addressline={}, 
            city={Xi'an},
%         citysep={}, % Uncomment if no comma needed between city and postcode
            postcode={710071}, 
            country={China}}
\author[1]{Chao Wang}[type=editor]
%\author[<aff no>]{<author name>}[<options>]
\author[2]{Junyuan Chen}[type=editor]
\author[2]{Jing Liu}[type=editor]
\cormark[2]
\cortext[2]{Corresponding author}
% Footnote of the second author

% Email id of the second author
\ead{neouma@mail.xidian.edu.cn}



\affiliation[2]{organization={Guangzhou Institute of Technology, Xidian University},
            %addressline={}, 
            city={Guangzhou},
%         citysep={}, % Uncomment if no comma needed between city and postcode
            postcode={510555}, 
            country={China}}



\begin{abstract}
Exact network structures and community partitions are valuable tools for modeling and analyzing complex systems. However, deriving them from the dynamics of complex systems is challenging.
Although many approaches have been devised to address network reconstruction (NR) and community detection (CD) independently, none of them consider explicit shareable knowledge across these two tasks. NR and CD from dynamics are natural synergistic tasks that motivate the proposed evolutionary multitasking NR and CD framework, called \emph{Network Collaborator}. In the method, the NR task explicitly shares the network structure with the CD task. And the CD task explicitly transfers the community partitions to assist the NR task. Moreover, to share knowledge from the NR task to the CD task, \emph{Network Collaborator} models the study of CD from dynamics to find communities in the dynamic network and then considers whether to transfer knowledge across tasks. To verify the performance of \emph{Network Collaborator}, we design a test suite of multitasking NR and CD problems based on various synthetic and real-world networks. The experimental results have demonstrated that joint NR with CD has a synergistic effect. Specifically, community partitions can be employed to improve the reconstruction accuracy of the network. Conversely, the network structure also better helps the discovery of community partitions. The code can be accessed at \url{https://github.com/xiaofangxd/EMTNRCD}.
\end{abstract}


\begin{highlights}
\item At the problem level, we propose joint optimization of NR and CD tasks from dynamics in a multitask setting. This is the first work that focuses on effectively transferring beneficial knowledge of these two tasks to achieve high accuracy.
\item We innovatively propose a solution to the multitasking network reconstruction and community detection problems, which existing evolutionary multitasking algorithms cannot solve directly.

\end{highlights}

% Keywords

\maketitle

% Main text
\section{Introduction}
Complex networks play an essential role in synchronizing and controlling complex dynamical systems \cite{strogatz2001exploring}, which have received extensive attention in many fields. Moreover, uncovering communities in networks complex offers coarse-graining relations between entities. However, in many natural complex systems, the network structure between entities and the complex systems’ nodal dynamics is unobservable. Instead, we may observe interdependent signals from the nodes in a complex network to infer these relationships. Thus, this paper focuses on two tasks: one for reconstructing network structure from dynamics; another one for discovering communities from dynamics.

\begin{figure}[t]
\centering
\includegraphics[width=0.48\textwidth]{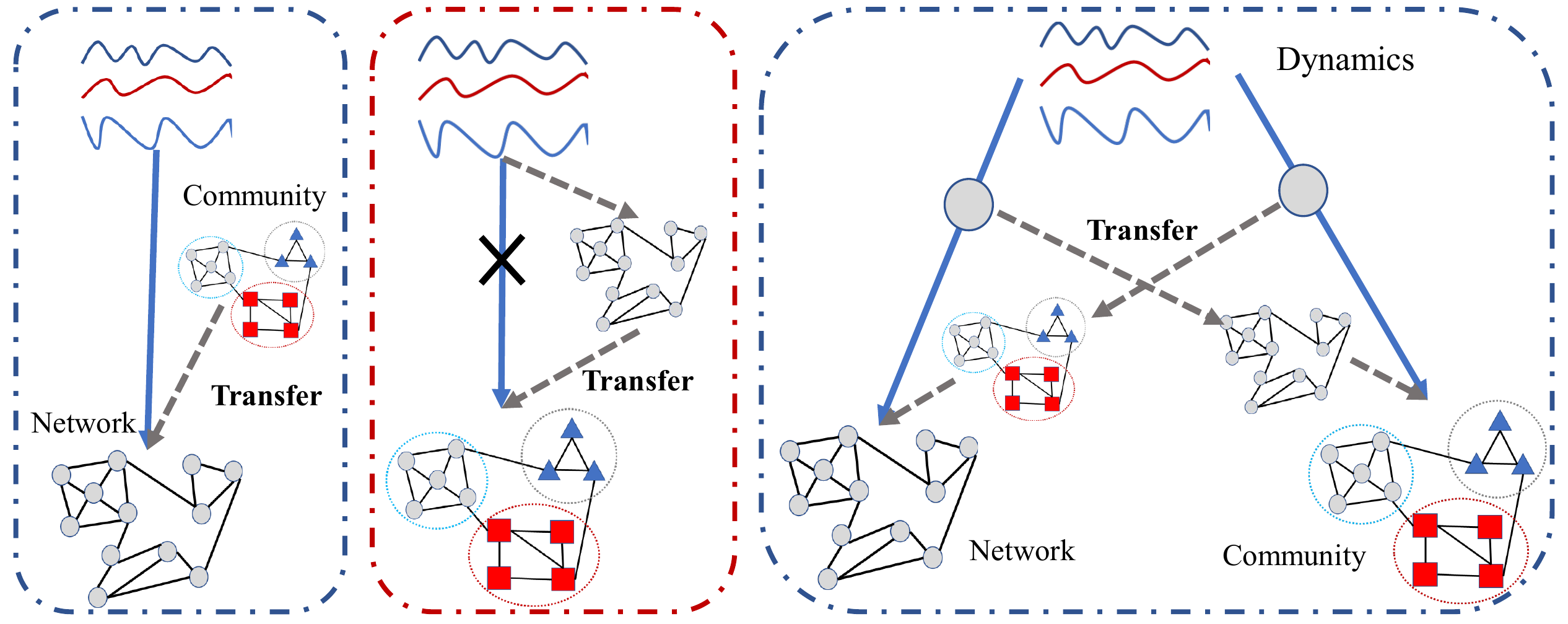}
\caption{The differences between our proposal and current methods. Left: CEMO-NR \cite{wu2020evolutionary} uses community partitions to aid the network reconstruction task. Middle: current methods first reconstruct the precise network structure from dynamics and then performs community detection on the obtained network. Right: the proposed multitasking network reconstruction and community detection problem.} 
%Note that no algorithm can detect communities directly from dynamics.
\label{fig1}
\vspace{-10pt}
\end{figure}
To solve the first task, a series of network reconstruction (NR) methods has been proposed to reconstruct network structure from dynamics, which are divided into two classes: model-free methods \cite{braunstein2008inference,margolin2006aracne,huynh2010inferring,marbach2012wisdom} and model-based methods \cite{han2015robust,wang2011network,nitzan2017revealing}. In model-free methods, the strength of a link among nodes is obtained by measuring the dependence from their dynamics in terms of correlations\cite{de2004discovery}, maximum entropy distributions \cite{braunstein2008inference}, mutual information \cite{margolin2006aracne}, random forest \cite{huynh2010inferring}, and ensemble method \cite{marbach2012wisdom}. In model-based methods, prior knowledge of the dynamics and interactions is provided. Then this knowledge is employed to infer the network structure, such as compressed sensing \cite{nitzan2017revealing,casadiego2017model}, evolutionary algorithm (EA) \cite{wu2016reconstructing,wu2019network,palafox2012reverse}, and online learning \cite{wu2020online,wu2020online1}.

Existing community detection (CD) methods are susceptible to specific design decisions when the network structure is unobservable. To address the challenges, most methods consist of two steps: first, select the NR methods to assess the similarity of any pair of factors in a complex system described above; second, convert the resemblance to a dense weighted network or a binary network \cite{chan2014decreased}. After determining the underlying network, the community detection methods are employed to discover clusters of the network, such as modularity \cite{newman2006modularity}, evolutionary algorithms (EAs) \cite{zhang2018network,folino2013evolutionary,gong2013complex,liu2020deep}, and deep learning \cite{liu2020deep}. These approaches assume that the edges have been observed accurately \cite{fortunato2010community}. 
However, these approaches do not consider knowledge transfer between NR and CD tasks, which may promote the accuracy of the two studies. We also find that a more precise network structure may promote the accuracy of community discovery, and better communities may promote the performance of the NR task. Thus, can the joint optimization of the NR and CD tasks obtain better performance?

This paper proposes an evolutionary multitasking NR and CD framework, termed \emph{Network Collaborator} (NC), to utilize the information of the network structure and community partition obtained from dynamics to improve each other. The phenomenon inspires this evolutionary multitasking framework that a more precise network structure can improve the accuracy of community discovery, and better communities can promote the performance of the NR task \cite{wu2020evolutionary}. Fig. \ref{fig1} shows the difference between our proposal and current methods. NC is inspired by evolutionary multitasking optimization \cite{gupta2016multifactorial}, a new paradigm for solving multiple tasks by taking advantage of the parallelism mechanism of evolutionary algorithms. Evolutionary multitasking optimization has been successfully applied to overcome many practical challenges in recent years because of its parallelism and easy scalability \cite{gupta2022half,tan2021evolutionary,wang2021learning}. These studies show that efficient knowledge transfer across tasks can improve the method's convergence characteristics\cite{feng2018evolutionary,wang2021solving}. Moreover, NR and CD tasks are often used to analyze complex systems simultaneously, which is a natural multitasking optimization problem. 
Thus, to share knowledge across the NR and CD tasks, we first establish a multitasking framework, where the community partition obtained from the CD task is explicitly transferred to improve the performance of the NR task, and the better network structures acquired from the NR task are explicitly transferred to enhance the performance of the CD task from dynamics.

However, existing evolutionary multitasking optimization methods cannot be employed to handle this problem because of the specificity of the transferred knowledge. Moreover, the gene coding of the NR task is continuous, and the gene coding of the CD task is discrete. Thus, we need to design a new evolutionary multitasking framework to handle this problem. Current EA-based CD methods assume that the network structure is known. Therefore they cannot directly address the CD task from dynamics.
Thus, we design a preprocessing stage to obtain the initial network structure before conducting the CD task. Moreover, communities are discovered from the given network in traditional CD tasks. In our framework, we transfer the new network structure obtained from the NR task to the CD task, which leads to the fact that the CD task differs from discovering communities from the static network. We first model this process as a dynamic CD problem that needs to infer communities from a dynamic network to overcome this issue. To assist the NR task, we design two local search strategies to utilize the inter-community and intra-community structural information transferred from the CD task. 

To validate the performance of NC, we design a test suite on four real-world networks and four synthetic networks by considering the evolutionary game (EG) model \cite{nowak1992evolutionary,szabo2007evolutionary} and resistor network (RN) model \cite{wang2011network,wu2016reconstructing}. Moreover, three state-of-the-art multiobjective EAs for the NR task and a state-of-the-art population-based CD algorithm of the dynamic network are embedded in NC. As shown in the experimental results, joining these two tasks has a synergistic effect, whereby the discovery of communities significantly increases the reconstruction accuracy, which in turn improves the performance of the CD task compared to performing these tasks in isolation.

\begin{table} [t]
\centering
\caption{The main symbols used in this paper.}
\label{tabmain}
\resizebox{.5\textwidth}{!}{ 
\begin{tabular}{l|l}
\hline
Parameters & Descriptions\\
\hline
$X$ &	The links between nodes in one network;\\
$N$ &	The number of nodes in one network;\\
$Y$ &	The observed data;\\
$P$ &	The $2 \times 2$ payoff matrices;\\
$S_i(t)$ &	The strategy of agent $i$ at the $t$th time;\\
$r_{ij}$	& The resistance of a resistor between nodes $i$ and $j$;\\
$V_i$ and $I_i$ &	The voltage and the total current at node $i$;\\
$C$ &	The community partition in one network;\\
$S$ &	The number of communities in $C$;\\
$l_s$ and $d_s$ &	\makecell[l]{The number of edges and the sum of the degree of \\ nodes in the $s$th community;}\\
$A$	& The confusion matrix of two community partitions;\\
$Z$ &	The multitasking NR and CD problems;\\
$N_1$ &	The population size for the NR task;\\
$N_2$ &	The population size for the CD task;\\
$TFE_1$ &	The number of function evaluations for the NR task;\\
$TFE_2$ &	The number of function evaluations for the CD task;\\
$\lambda$ &	\makecell[l]{The share of the $TFE$ used for the normal \\ optimization stage in NC;}\\
$t_1$ &	\makecell[l]{The number of function evaluations for the knowledge \\ transfer from the CD task to NR task in NC;}\\
$P_{NR}^{(t)}$ &	The population for the NR task in the $t$th generation;\\
$X(t)$ &	\makecell[l]{The network structure obtained from the NR task \\ in the  $t$th generation;}\\
$C^{(t)}$ &	\makecell[l]{The community partition obtained from the CD task \\ in  the $t$th generation;}\\
$S{(t)}$ &	The number of communities in $C^{(t)}$;\\
$D$ &	The number of decision variables for the NR task;\\
$Li$ &	The number of links in the network;\\
$S_c$ &	The number of communities in real networks;\\
$N_s-L$ &	$N_s$ response sequences with $L$ time points each;\\
\hline
\end{tabular}}
\end{table}

The contributions of the NC are summarized as follows:
\begin{enumerate}
    \item %At the problem level, we first propose joint optimization of NR and CD tasks from dynamics in a multitask setting. There are community-assisted EA-based network reconstruction schemes and network reconstruction-assisted EA-based community detection schemes. However, they are all one-way assists and do not consider two-way assists to achieve each other. From the problem level, this is the first work that focuses on effectively transferring beneficial knowledge of these two tasks to achieve high accuracy.
    The current network reconstruction and community detection tasks are handled separately. We found that the network reconstruction and community detection tasks can facilitate each other. The knowledge that can be transferred between the two tasks is the network structure in the network reconstruction task and the community structure in the community detection task.
    \item  %From a technical point of view, we innovatively propose a solution to the multitasking network reconstruction and community detection problems, which existing evolutionary multitasking algorithms cannot solve directly. For the evolutionary multitasking community, this work expands the application scope of evolutionary multitasking optimization and makes technical innovations for this problem.
    We design a multitask community detection and network reconstruction framework to exploit the discovered knowledge for transfer between the two tasks, overcoming the shortcomings of current evolutionary multitask optimization frameworks that cannot use this knowledge.
\end{enumerate}

The rest of this paper is organized as follows. Section 2 reviews EA-based NR methods, EA-based CD methods, and evolutionary multiobjective multitasking optimization. In Section 3, we introduce the problem formulation. The details of NC are shown in Section 4. Section 5 gives the experimental results of the designed test suite to illustrate the effectiveness of our methods. Finally, Section 6 summarizes the work in this paper and discusses the potential directions.

\section{Related Work}
\subsection{EA-based NR Methods}
The NR task aims to reconstruct the links between each pair of entities. Various EA-based methods were proposed to overcome NR problems. Here, several traditional methods are introduced to show our motivation for NC. These approaches have a similar inference pattern in handling the task of NR from dynamics. The inference model, such as fuzzy cognitive maps \cite{wu2020online,wu2020time,wu2017learning}, S-system \cite{kimura2005inference,palafox2012reverse}, recurrent neural network \cite{xu2007inference}, is used to model the observed data obtained from dynamics. Then the EA is employed to optimize the parameters of the inference model. Finally, the designed inference model is used to obtain the network structure of the complex system. More accurate models and high-performance optimizers are the keys to these approaches. Several approaches are proposed to reconstruct the network structure based on the complex behavior \cite{wu2016reconstructing,wu2019network}. However, these methods do not consider the influence of community partition on the NR task, which may improve the NR task’s performance, especially for the high-dimensional NR problems. The most related work (CEMO-NR) proposed in \cite{wu2020evolutionary} employed community partition information to aid the NR task by decomposing the original problem into several low-dimensional subproblems. However, CEMO-NR does not consider the CD task nor transfer helpful knowledge to improve the performance of the CD task performed alone. Moreover, both tasks were not optimized simultaneously.

\subsection{EA-based CD Methods}
A series of EA-based approaches have been proposed to discover the community partition from the different networks \cite{pizzuti2017evolutionary}, such as large-scale networks \cite{zhang2018network,gong2013complex}, dynamic networks \cite{folino2013evolutionary,zeng2019consensus,10071529}, attributed networks \cite{teng2019overlapping}, multilayer Network\cite{10102396,9802693,10035921}, and signed social networks \cite{liu2014multiobjective}.
Moreover, several works are proposed to handle the more difficult task of discovering overlapping communities \cite{wen2016maximal,zhang2017mixed,teng2019overlapping}.
However, they are not able to discover communities from dynamics. These works assume that the exact network structure has been obtained before performing the CD. Our proposed NC is the first EA-based framework to overcome this challenge.

In general, most methods need to perform the following two steps to infer communities from dynamics: First, one NR method is chosen to reconstruct network structure from dynamics. Second, the EA-based CD methods are employed to detect network clusters obtained by the first step. This process is described in Fig. \ref{fig1}. This type of approach assumpts that the edges are accurately observed. Meanwhile, the error suffered by the CD will enlarge the error of the whole process. Unlike the above methods, Hoffmann \emph{et al.} \cite{hoffmann2020community} proposed a Bayesian hierarchical model to directly discover communities from time series without considering the process of the NR. However, neither of these approaches attempts to perform the NR together with the CD, and none of them considers \textit{knowledge transfer} among these two tasks, which may promote the accuracy of the CD from dynamics.

Using the NR and CD tasks characteristics, the proposed NC employs the community partition obtained by the CD task to improve reconstruction accuracy and operates the better network obtained by the NR task to find a better community partition.

\subsection{Evolutionary Multiobjective Multitasking Optimization}
The NR and CD tasks are modeled as multiobjective multitasking optimization problems in this paper. Therefore, we review the existing methods. Inspired by the multifactorial inheritance across organisms and the parallelism of population-based search, a multiobjective multifactorial evolutionary algorithm (MO-MFEA) was proposed in \cite{gupta2016multiobjective} to optimize multiple tasks simultaneously in a single population. MO-MFEA designed a fixed-parameter rmp to control the knowledge transfer simply, which does not consider the relationship between tasks. Recently, Bali \emph{et al.} \cite{bali2020cognizant} introduced MO-MFEA-II to overcome this issue. In this method, an online transfer parameter estimation maintained a transfer parameter matrix RMP to adaptively control the degree of knowledge. Besides, some methods with adaptive knowledge transfer capabilities have also attracted great attention. An explicit multiobjective multitasking evolutionary algorithm is proposed by Feng \emph{et al.} \cite{feng2018evolutionary}, in which each task is assigned an independent solver. Then for every two tasks, a denoising autoencoder is designed to learn linear mappings across tasks. Lin \emph{et al.} \cite{lin2019multiobjective} introduced a novel evolutionary multiobjective multitasking framework based on incremental Naive Bayes classifiers to find helpful knowledge (solutions) during the multitasking search. To improve the convergence rate, Liang \emph{et al.} \cite{9817402} used generative strategies based on generative adversarial networks and inertial differential evolution to produce transferable knowledge and high-quality offspring. Recently, Wang \emph{et al.} \cite{9721409} proposed a novel dual-neighborhood evolutionary algorithm, where the neighborhood is viewed as a bridge to enable efficient knowledge transfer between different multiobjective optimization tasks. Existing methods mainly focus on continuous optimization, which cannot be directly employed in NR and CD tasks with discrete encoding.

Recently, some evolutionary multitasking methods have been proposed to solve NR or CD problems. Wu \emph{et al.}  \cite{wu2021evolutionary} proposed an evolutionary multitasking framework for multilayer NR, which exploits the shared knowledge across component layers to improve reconstruction performance. Lyu \emph{et al.} \cite{9802693} presented a CD method for multilayer networks based on evolutionary multitasking optimization and evolutionary clustering ensemble. This approach can divide specific communities for each component layer and composite communities shared by all layers. The above methods improve the performance of NR or CD tasks through knowledge transfer between component layers. However, the shared knowledge between NR and CD is ignored.

\section{Problem Formulation}
The main symbols of this paper are summarized in Table \ref{tabmain}. This section introduces the problem formulation of multiobjective multitasking optimization, NR problems, and CD problems. Then we discuss how to model NR problems and CD problems as multitasking NR and CD.

\subsection{Multiobjective Multitasking Optimization}
For $K$ minimization tasks, the multiobjective multitasking optimization problem is mathematically formulated as follows:
\begin{equation}
\left\{ \begin{matrix}
  \mathop {\min }\limits_{{X_i}} {{\bf{F}}_i}\left( {{{\bf{x}}_i}} \right) = \left( {f_i^1\left( {{{\bf{x}}_i}} \right),f_i^2\left( {{{\bf{x}}_i}} \right), \ldots ,f_i^{{m_i}}\left( {{{\bf{x}}_i}} \right)} \right) \hfill \cr
  s.t.\;{{\bf{x}}_i} = [x_i^1,x_i^2, \ldots ,x_i^{{n_i}}] \in {D_i},i = 1,2, \ldots ,K \hfill \cr\end{matrix}  \right.
\end{equation}
where ${\bf{F}}_i({\bf{x}}_i)$ is the $i$-th multiobjective optimization task and $D_i$ is the search space for optimization task $i$. $n_i$ and $m_i$ are the number of objective functions and the dimensionality of $x_i$ in the $k$-th task, respectively. Suppose that $x_i^{(1)}$ and $x_i^{(2)}$ be two solutions for the $i$-th multiobjective optimization task, $x_i^{(1)}$ is said to Pareto dominate $x_i^{(2)}$, if and only if $f_i^j(x_i^{(1)})\leq f_i^j(x_i^{(2)}),\forall j\in\{1, 2, \cdots, m_i\}$ and there exists at least one objective $f_i^k (k\in\{1, 2, \cdots, m_i\})$ satisfying $f_i^j(x_i^{(1)})< f_i^j(x_i^{(2)})$. $x_i^*$ is Pareto optimal if there is no $x_i$ such that $x_i$ dominates $x_i^*$. The set of all Pareto optimal solutions is named the Pareto set. The projection of the Pareto set in the objective space is called the Pareto front. In summary, multiobjective multitasking optimization aims to find a set of solutions approximating the Pareto front for each multiobjective optimization task, which should be as close as possible to the Pareto front and distributed evenly and widely over the Pareto front \cite{bali2020cognizant}.

% By exploiting the potential similarities between multiple tasks, MMTO aims to improve the performance of each multiobjective optimization task in terms of convergence and diversity. The core of designing the EMMTO algorithm is two issues \cite{lin2019multiobjective}: \textit{For each task, 1) how to select practical transferred knowledge from other tasks? and 2) how to make full use of the shared knowledge from other tasks?}

\subsection{Network Reconstruction}
A network can be regarded as a graph $G = (V, E)$, where the nodes (vertices) in \textit{V} represent the concepts, and the edges in \textit{E} represent the relationship among individuals \cite{newman2003structure}. $V = \{v_1, v_2, \cdots, v_N\}$ is a set of \textit{N} nodes and $E = \{(v_i, v_j) | v_i, v_j \in V, i\neq j \}$ for undirected networks. The network structure \textit{X} between nodes is defined as follows:
\begin{equation}
   X = \left[ {\begin{array}{*{20}{c}}
{{x_{11}}}& \cdots &{{x_{1N}}}\\
 \vdots & \ddots & \vdots \\
{{x_{N1}}}& \cdots &{{x_{NN}}}
\end{array}} \right]
\end{equation}
where $x_{ij}\in\{0, 1\}$ is the connection across nodes \textit{i} and \textit{j} and \textit{N} represents the number of nodes.
Let \textit{Y} and \textit{h}(\textit{X}, \textit{Y}) be the observed data and the inference model simulation from the candidate network structure, respectively. The goal of the NR is to infer the connections across each pair of nodes according to \textit{Y} and \textit{h}(\textit{X}, \textit{Y}). In general, the NR problem can be expressed as follows: \cite{newman2003structure}:
\begin{equation}
\begin{array}{l}
\mathop {\min }\limits_X F\left( X \right) = \left( {h\left( {X,Y} \right),g(X)} \right)\\
s.t.\;X \in {\left\{ {0,1} \right\}^{N \times N}}
\end{array}
\end{equation}
The definition of \textit{h}(\textit{X}, \textit{Y}) and \textit{g}(\textit{X}) is determined by the complex network’s inference model. The detailed model can be found in \cite{wu2021evolutionary}. We introduce two common NR problems in Appendix.

\subsection{Community Detection}
It is difficult to handle the task of detecting communities from dynamics. We also introduce the traditional CD from network methods, which will be employed in our framework. 
CD aims to divide all network nodes into multiple communities with dense intra-community links and sparse inter-community links \cite{newman2006modularity}.  Let $C = \{C_i | C_i \subseteq V, C_i \neq \varnothing, i=1, 2, \cdots, S\}$ be a set of $S$ communities obtained from $G$. Thus, it must satisfy the following conditions:
\begin{equation} \label{eq4}
\begin{array}{l}
\bigcup\limits_{i = 1}^S {{C_i}}  = V\\
\forall i \ne j,\;{C_i} \ne {C_j},i,j \in \{ 1,2, \ldots ,S\} 
\end{array}
\end{equation}

For CD problems, modularity \textit{Q} is one of the most well-known functions for evaluating the quality of network partitions, which can be expressed as follows:
\begin{equation} \label{eq5}
    Q = \sum\limits_{s = 1}^S {\left[ {\frac{{{l_s}}}{e} - {{\left( {\frac{{{d_s}}}{{2e}}} \right)}^2}} \right]}
\end{equation}
where \textit{e} denotes the total number of edges; $l_s$ and $d_s$ represent the number of edges and the sum of the degree of nodes in the \textit{s}-th community, respectively. The larger the \textit{Q} value, the better the CD results of the network.

Normalized mutual information (NMI) is a commonly used function to measure the similarity of two community partitions. Let $B_1$ and $B_2$ be two partitions of a network and \textit{A} be a confusion matrix, where element $A_{ij}$ is the number of nodes that belong to both the \textit{i}-th community of $B_1$ and the \textit{j}-th community of $B_2$. Then NMI can be expressed as follows:
\begin{equation} \label{eq6}
    {\rm N}{\rm M}{\rm I}({B_1},{B_2}) = \frac{{ - 2\sum\limits_{i = 1}^{{S_1}} {\sum\limits_{j = 1}^{{S_2}} {{A_{ij}}\log \left( {\frac{{{A_{ij}} \cdot N}}{{{A_{i \cdot }} \cdot {A_{ \cdot j}}}}} \right)} } }}{{\sum\limits_{i = 1}^{{S_1}} {{A_{i \cdot }}\log \left( {\frac{{{A_{i \cdot }}}}{N}} \right) + } \sum\limits_{i = 1}^{{S_2}} {{A_{ \cdot j}}\log \left( {{A_{ \cdot j}} \cdot N} \right)} }}
\end{equation}
where $S_1$ and $S_2$ are number of communities in $B_1$ and $B_2$, respectively; $A_{i\cdot}$ and $A_{\cdot j}$ are the sum of the elements of \textit{A} in the \textit{i}-th row and \textit{j}-th column, respectively. Generally speaking, the larger the NMI value, the more similar $B_1$ and $B_2$.

\subsection{Multitasking Network Reconstruction and Community Detection}

\begin{figure*}[htbp]
\centering
\includegraphics[width=0.8\textwidth]{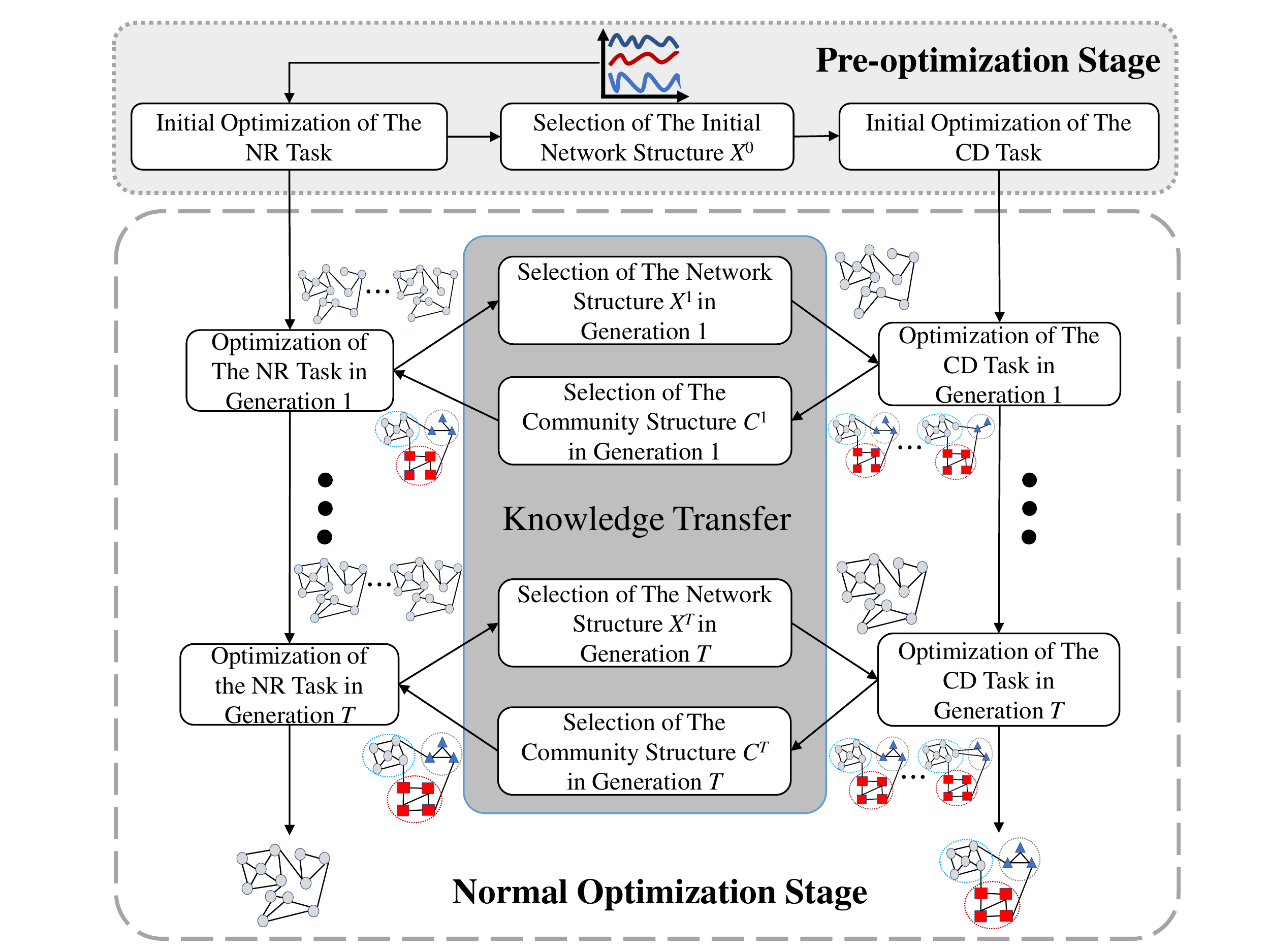}
\caption{Outline of the proposed method consisting of two different optimization stages: pre-optimization and normal optimization.}\label{fig3}
\end{figure*}

The NR and CD tasks are two hot topics in researching complex network systems. The former aims to infer the network structure from the limited observed data, while the latter aims to explore the complex interactions and relationships among nodes from dynamics. As is shown in Fig. \ref{fig1}, we can generally reconstruct the network structure from dynamics first and then detect the community partition.

The work proposed in \cite{blondel2008fast} shows that social networks naturally tend to cluster into groups or communities. The nodes in the same group link more densely than nodes outside the group. Inspired by this, the explicit \textit{knowledge transfer} across the CD task and NR task is considered to improve each task’s performance, modeled as a multiobjective multitasking optimization problem, as shown in Fig. \ref{fig1}. In the optimization process, the community partition obtained from the CD task can be considered useful knowledge to reconstruct the connection between nodes. Similarly, the network structure obtained from the NR task can also be transferred to assist the CD task. The detailed procedure is shown in Section IV.

%\begin{figure} [htb]
%\centering
%\includegraphics[width=0.5\textwidth]{fig2.pdf}
%\caption{A process of multitasking NR and CD.} 
%\label{fig2}
%\end{figure}

\section{Proposed Method}
This section introduces the outline of the proposed method first. Then the explicit knowledge transfer from the NR task to the CD task is presented. Next, we also give the optimization process of the NR task, including the knowledge transfer from the CD task to the NR task. Finally, the framework of the proposed method is presented.

\subsection{Overview}

The goal of NC is to simultaneously reconstruct the network structure and detect the community partition from the given limited observed data by transferring knowledge across the NR and CD tasks. The outline of the proposed method is shown in Fig. \ref{fig3}, consisting of two different optimization stages: pre-optimization and normal optimization. In the pre-optimization stage, we first optimize the NR task with a fixed number of function evaluations by employing the general population-based NR algorithm. Then, the initialized network structure is selected from the optimized population. Next, based on the initial network structure, we optimize the CD task with a fixed number of function evaluations by employing the general population-based CD algorithm. After the pre-optimization stage, we obtain the initial network and community partitions used in the subsequent stages. The NR and CD tasks are optimized simultaneously with explicit knowledge transfer in the normal optimization stage. In optimizing each generation of the CD task, the network structures obtained from the NR task in the current generation and the previous generation can be considered the snapshots of a dynamic network at two consecutive time steps, viewed as prior knowledge to assist the CD task. Thus, we model this process as the dynamic CD problem and then solve the CD task using the general population-based CD algorithm of a dynamic network. In optimizing each generation of the NR task, the same population-based NR algorithm as the pre-optimization stage is employed. Since the nodes in the same community link more densely than nodes outside the community, the community partition obtained from the CD task is transferred to help reconstruct the links between nodes in the NR task, which is inspired by the work \cite{wu2020evolutionary}. Next, we introduce the procedure of explicit knowledge transfer across the NR task and CD task.

\subsection{Knowledge Transfer from the NR Task to CD Task}

We usually need prior knowledge about the network structure $X$ when solving the CD task. In the evolutionary process of NC, more and more precise network structures are obtained with the development of the optimization of the NR task. Inspired by this, the network structures obtained from the NR task at consecutive generations can be viewed as multiple snapshots of a dynamic network to assist the CD task. To realize the above ideas, the following two key issues need to be resolved: 1) How to obtain the network structure $X^{(t)}$ from optimized population $P_{NR}^{(t)}$ of the NR task in the $t$-th generation? 2) How to take full advantage of the network structure $X^{(t)}$ to assist the CD task?

\textit{1) How to obtain the network structure $X(t)$ from optimized population $P_{NR}^{(t)}$ of the NR task in the \textit{t}-th generation?}

How to select the network structure $X^{(t)}$ has a significant effect on the performance of NC. The NR task aims to obtain a set of Pareto-optimal network structures, which represent a trade-off between multiple objectives such as reconstruction performance and network sparsity. Unlike single-objective optimization, the unique best network structure cannot be directly obtained from $P_{NR}^{(t)}$. In this paper, network structure $X(t)$ is selected by applying the crowding distance metric to the first nondominated front of the population $P_{NR}^{(t)}$ reported in \cite{deb2002fast}. The pseudo-code of the selection operator is given in Algorithm \ref{alg1}. Firstly, the fast nondominated sorting method \cite{deb2002fast} is employed to partition $P_{NR}^{(t)}$ into \textit{L} nondominated fronts ${F_1, \cdots, F_L}$, where $F_1$ is the first nondominated front, and $F_L$ is the last one. The network structures in $F_1$ are individuals with good convergence in $P_{NR}^{(t)}$. Next, the crowding distance of each network structure in $F_1$ is calculated. The network structure $X_{(t)}$ in $F_1$ with the largest crowding distance value is selected to assist the CD task. This unique network structure is far away from other network structures in $F_1$. If there are two extreme network structures in $F_1$, we choose one at random. Therefore, the selection process comprehensively considers the diversity and convergence of the network structure set.

\begin{algorithm}[t]
 \caption{Selection} 
 \label{alg1}
 \begin{algorithmic}[1]
  \REQUIRE $P_{NR}^{(t)}$: Optimized population of the NR task in the \textit{t}-th generation.
  \ENSURE Output $X^{(t)}$: Network structure obtained from the NR task in the \textit{t}-th generation.
  \STATE $\{F^1, ..., F^L\}$ $\leftarrow$ \textit{Fast Nondominated Sorting}  ($P_{NR}^{(t)}$);
  \STATE count $\leftarrow$ Count the number of network structures in $F^1$;
  \IF{$count > 2$}
  \STATE $X^{(t)} \leftarrow{\arg \mathop {\max }\limits_{X \in {F^1}} Crowding Distance({F^1})}$;
  \ELSE{}
  \STATE $X^{(t)}$ $\leftarrow$ Randomly choose a network structure from $F^1$;
  \ENDIF
%  \RETURN $X^{(t)}$
 \end{algorithmic}
\end{algorithm}

\textit{2) How to take full advantage of the network structure $X^{(t)}$ to assist the CD task?}

After obtaining the network structure $X^{(t)}$ from the optimized population $P_{NR}^{(t)}$, we need to consider how to apply the information of network structure $X^{(t)}$ to assist the CD task. The evolution analysis of the community partition of dynamic networks is also one of the current research hotspots in network sciences. In the evolutionary process of NC, the network structures $X^{(t)}$ and $X^{(t-1)}$ obtained from the NR task for two consecutive generations can be regarded as the snapshots of a dynamic network \textit{G} in two consecutive time steps. For convenience, a dynamic network \textit{G} is denoted as $G = \{G^{(t)}= (V, E^{(t)}), t=0, 1, \cdots, T\}$, where $G^{(t)}$ represents a snapshot of \textit{G} in the \textit{t}-th generation of NC. Let $C^{(t)} = \{C_j^{(t)} | C_j^{(t)} \subseteq V, C_j^{(t)} \neq \varnothing, j=1, 2, \cdots, S^{(t)}\}$ be the community partition containing $S^{(t)}$ communities obtained from $G^{(t)}$ and $C_i^{(t)} \cap C_j^{(t)} = \varnothing $ for any $C_i^{(t)}$, $C_j^{(t)} \in C$. Then, the CD in dynamic networks aims to find the communities $C = \{C^{(0)}, C^{(1)}, \cdots, C^{(T)}\}$. To uncover dynamic networks’ evolutionary behavior, a framework named temporal smoothness is proposed by Chakrabarti \emph{et al.} \cite{chakrabarti2006evolutionary}, emphasizing that the network should not shift significantly from one timestep to the next. A cost function $f^{(t)}$ combining snapshot cost $f_1^{(t)}$ and temporal cost $f_2^{(t)}$ at generation \textit{t} is described as follows:
%\begin{equation}
   % {f^{\left( t \right)}}\left( {{C^{\left( t \right)}}} \right) = \left\{ \begin{array}{l}
%f_1^{\left( t \right)}\left( {{C^{\left( t \right)}}} \right),t = 0\\
%\alpha  \cdot f_1^{\left( t \right)}\left( {{C^{\left( t \right)}}} \right) + \left( {1 - \alpha } \right) \cdot f_2^{\left( t \right)}\left( {{C^{\left( t \right)}},{C^{\left( {t - 1} \right)}}} \right),t = 1, \ldots ,T
%\end{array} \right.
%\end{equation}
\begin{equation}\label{eq7}
\left\{ \begin{array}{l}
f_1^{\left( t \right)}\left( {{C^{\left( t \right)}}} \right),t = 0\\
\alpha  \cdot f_1^{\left( t \right)}\left( {{C^{\left( t \right)}}} \right) + \left( {1 - \alpha } \right) \cdot f_2^{\left( t \right)}\left( {{C^{\left( t \right)}},{C^{\left( {t - 1} \right)}}} \right),t = [1, T]
\end{array} \right.
\end{equation}
where $ \alpha \in [0, 1]$ is a balance parameter that controls the trade-off between $f_1^{(t)}$ and $f_2^{(t)}$. The first term $f_1^{(t)}$ evaluates the quality of the community partition at generation \textit{t}, and the second term $f_2^{(t)}$ evaluates the similarity of community partitions between generation \textit{t} and \textit{t}-1. To automatically obtain the best trade-off between the above two items, Folino \emph{et al.} \cite{folino2013evolutionary} transformed the CD in dynamic networks into a multiobjective optimization problem, which can be expressed as follows:
\begin{equation} \label{eq8}
    \left\{ \begin{array}{l}
\mathop {\max }\limits_{{C^{\left( t \right)}}} f_1^{\left( t \right)}\left( {{C^{\left( t \right)}}} \right),t = 0\\
\mathop {\max }\limits_{{C^{\left( t \right)}}} \left( {f_1^{\left( t \right)}\left( {{C^{\left( t \right)}}} \right),f_2^{\left( t \right)}\left( {{C^{\left( t \right)}},{C^{\left( {t - 1} \right)}}} \right)} \right),t = 1,\ldots ,T
\end{array} \right.
\end{equation}

Based on the reported results in \cite{liu2019evolutionary,liu2020detecting}, modularity \textit{Q} in Eq. \ref{eq5} and NMI in Eq. \ref{eq6} are used as the two optimization objectives for finding communities of dynamic networks in this paper. The former evaluates the quality of community partitions, while the latter evaluates the similarity of community partitions in two consecutive generations. Then we can optimize Eq. \ref{eq8} by the general population-based CD algorithm of the dynamic network to obtain communities $C = \{C^{(0)}, C^{(1)}, \cdots, C^{(T)}\}$. Based on Eq. \ref{eq8}, the CD task of a dynamic network composed of the network structure obtained by two consecutive generations \textit{t} and \textit{t}-1 in this paper is shown as follows:
\begin{equation} \label{eq9}
\begin{array}{*{20}{c}}
{\mathop {\max }\limits_{{C^{\left( t \right)}}} \left( {Q\left( {{C^{\left( t \right)}}} \right),NMI\left( {{C^{\left( t \right)}},{C^{\left( {t - 1} \right)}}} \right)} \right)}\\
\begin{array}{l}
s.t.\;{C^{\left( t \right)}}{\rm{ = }}\left\{ {C_j^{\left( t \right)}\left| {C_j^{\left( t \right)} \in V,} \right.C_j^{\left( t \right)} \ne \emptyset ,j = 1,2,....,{S^{\left( t \right)}}} \right\}\\
\forall C_i^{\left( t \right)},C_j^{\left( t \right)} \in {C^{\left( t \right)}},C_i^{\left( t \right)} \cap C_j^{\left( t \right)} = \emptyset 
\end{array}
\end{array}
\end{equation}
\subsection{Knowledge Transfer from the CD Task to NR Task}

\begin{algorithm}[t]
\small
 \caption{Knowledge Transfer from the CD Task to NR Task} 
 \label{alg2}
 \begin{algorithmic}[1]
  \REQUIRE $N_1$: Population size for the NR task, $P_{NR}^{(t)}$: Population obtained from the normal optimization stage of the NR task in \textit{t}-th generation,$C^{(t)}=\{C_j^{(t)} | C_j^{(t)} \subseteq V, C_j^{(t)} \neq \varnothing, j=1, 2, \cdots, S^{(t)}\}$: Network structure obtained from the NR task in the \textit{t}-th generation, $\alpha$: Population size for the knowledge transfer from the CD task to NR task, $t_1$: Number of function evaluations for the knowledge transfer from the CD task to NR task.
  \ENSURE Output $P_{NR}^{(t)}$: Population obtained from the CD task to NR task.
  \STATE $ X'\leftarrow Selection (P_{NR}^{(t)})$; // Algorithm 1;
  \STATE // local search operator $ls_1$ on $X'$ by utilizing the structural information within a community.
  \STATE $C'\leftarrow $Randomly choose a community from $C^{(t)}$;
  \STATE $P_{in} \leftarrow Initialization_{within} (\alpha, X', C')$;
  \WHILE{the used number of function evaluations$ \leq t_1$}
  \STATE $O\leftarrow Offspring Generation_{within} (P_{in})$;
  \STATE $P_{in}\leftarrow Environment Selection (\{P_{in}, O\}, \alpha)$;
  \ENDWHILE
  \STATE // local search operator $ls_2$ on $X'$ by utilizing the structural information between communities.
  \STATE $P_{out} \leftarrow Initialization_{between} (\alpha, X', C^{(t)})$;
  \WHILE{the used number of function evaluations$ \leq t_1$}
  \STATE $O\leftarrow Offspring Generation_{within} (P_{out})$;
  \STATE $P_{out}\leftarrow Environment Selection (\{P_{out}, O\}, \alpha)$;
  \ENDWHILE
  \STATE $P_{NR}^{(t)}\leftarrow Environment Selection (\{P_{NR}^{(t)}, P_{in}, P_{out}\}, N_1)$;
 % \RETURN $P_{NR}^{(t)}$
 \end{algorithmic}
\end{algorithm}

\begin{figure*} [t]
\centering
\includegraphics[width=0.65\textwidth]{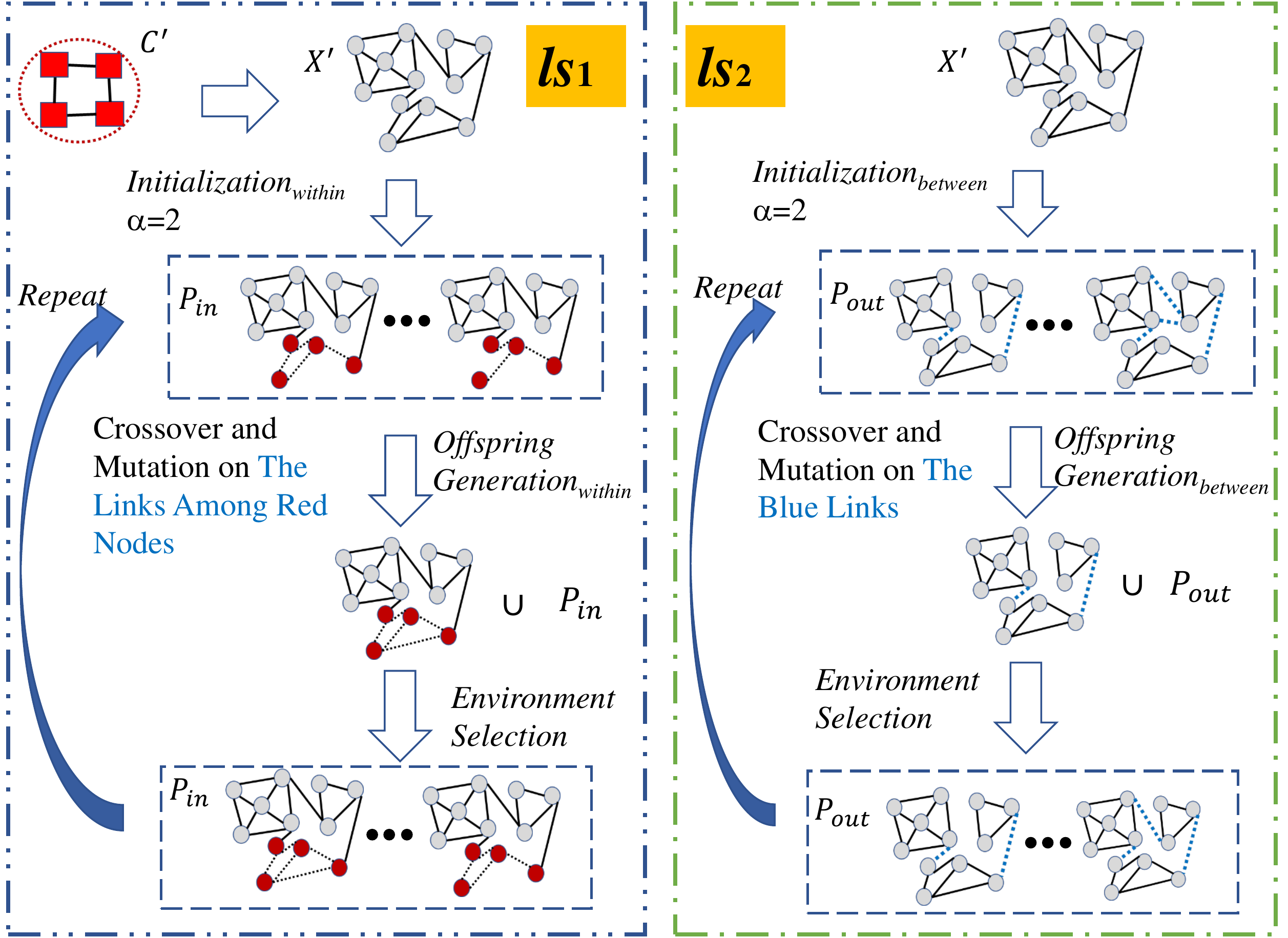}
\caption{Outline of two local search operators in knowledge transfer from the CD task to NR task.} \label{fig4}
\end{figure*}

There may be multiple communities in the real network, and the connections between nodes in one community are generally tight. Inspired by this phenomenon, the community partition $C^{(t)}$ from the CD task in the \textit{t}-th generation is explicitly transferred to the NR task to take advantage of the internal tightness of the community. The knowledge transfer from the CD task to the NR task is proposed to realize the above ideas, as shown in Algorithm \ref{alg2}. Firstly, the solution $X^{
'}$ is obtained from the $P_{NR}^{(t)}$ by the same selection method as algorithm \ref{alg1}. Then we perform two local searches $ls_1$ and $ls_2$ on $X^{'}$ by utilizing the link information within and between communities, respectively. Further, the environment selection procedure that is the same as the procedure of the population-based NR algorithm in the normal optimization stage is performed on the union set ${P_{NR}^{(t)}, P_{in}, P_{out}}$ to obtain the new population $P_{NR}^{(t)}$ with $N_1$ chromosomes after knowledge transfer. Next, we introduce how to perform two local search operators in detail. Moreover, the example of these two operators can be found in Fig. \ref{fig4}.

A community $C^{'}$ is randomly selected from $C^{(t)}$, and then a local search operator $ls_1$ (lines 3-8 in Algorithm \ref{alg2}) is proposed by utilizing the links information within a community $C^{'}$. Firstly, each link relationship within the community $C^{'}$ in the network $X^{'}$ is changed with a probability of 0.8, while other link relationships remain unchanged. Then an initial population $P_{in}$ with $\alpha$ chromosomes can be obtained by repeating this operator (line 4 in Algorithm \ref{alg2}). In the main loop, $\alpha$ individuals are generated by repeating the following operation. Two solutions from the population $P_{in}$ are selected randomly, and then the single-point crossover and bitwise mutation operator are performed on the decision variables corresponding to the link relationship within the community $C^{'}$ (line 6 in Algorithm \ref{alg2}). Finally, the environment selection procedure that is the same as the population-based NR algorithm procedure in the normal optimization stage is used to select solutions that survive to the next generation (line 7 in Algorithm \ref{alg2}).

Besides, a local search operator $ls_{2}$ (lines 10-14 in Algorithm \ref{alg2}) is proposed by utilizing the links information between communities, and its difference from $ls_{1}$ includes two operations: Initialization and offspring generation. We only need to modify the link information within a community $C^{'}$ used in the above two operators of ls1 to the link information between communities to obtain $ls_2$. In summary, as prior knowledge, the link information within and between communities is embedded in two local search operators to optimize the NR task.

\subsection{Framework of Network Collaborator}

\begin{algorithm}[t]
\small
 \caption{Network Collaborator}
 \label{alg3}
 \begin{algorithmic}[1]
  \REQUIRE \textit{Z}: MNRCDPs; $N_1$: Population size for the NR task; $N_2$: Population size for the CD task; $TFE_1$: Number of function evaluations for the NR task; $TFE_2$: Number of function evaluations for the CD task; $\lambda, \alpha, t_1, t_2$: Key parameters; $OptimizerNR$: Optimizer for the NR task; $OptimizerCD_{pre}$: Optimizer for the CD task in pre-optimization stage; $OptimizerCD_{normal}$: Optimizer for the CD task in normal optimization stage.
  \ENSURE Output learned network structures $X^*$ and community partition $C^*$.
  \STATE $P_{NR}^{(0)} \leftarrow Initialization (N_1, Z)$;
  \STATE // pre-optimization stage of the NR task
  \STATE $P_{NR}^{(0)} \leftarrow OptimizerNR (N_1, Z, P_{NR}^{(0)}, (1-\lambda)\times TFE_1)$;
  \STATE $X^{(0)} \leftarrow Selection (P_{NR}^{(0)})$; // \textbf{Algorithm \ref{alg1}}
  \STATE // pre-optimization stage of the CD task
  \STATE $C^{(0)} \leftarrow OptimizerCD_{pre} (N_2, Z, (1-\lambda)\times TFE_2)$;
  \STATE $t\leftarrow 1$;
  \WHILE{termination criterion not fulfilled}
  \IF{used $FE_1\leq \lambda \times TFE_1$}
  \STATE // normal optimization stage of the NR task
  \STATE $P_{NR}^{(t)} \leftarrow OptimizerNR (N_1, Z, P_{NR}^{(t)}, N_1)$;
  \IF{used $FE_2\leq \lambda \times TFE_2$}
  \STATE $X^{(t)} \leftarrow Selection (P_{NR}^{(t)})$; // \textbf{Algorithm \ref{alg1}}
  \STATE // normal optimization stage of the CD task
  \STATE $C^{(t)} \leftarrow OptimizerCD_{normal} (N_2, Z, X^{(t)}, C^{(t-1)}, t_2)$;
  \STATE $P_{NR}^{(t)} \leftarrow KTransfer_{CD\rightarrow NR} (N_1, P_{NR}^{(t)}, C^{(t)}, \alpha, t_1)$; // \textbf{Algorithm \ref{alg2}}
  \ENDIF
  \ENDIF
  \STATE $t\leftarrow t+1$;
  \ENDWHILE
  \STATE $X^{*} \leftarrow Selection (P_{NR}^{(t)})$; // \textbf{Algorithm \ref{alg1}}
  \STATE $C^{*}\leftarrow C^{t}$
%  \RETURN $X^{*}$ and $C^{*}$
 \end{algorithmic}
\end{algorithm}

The general framework of NC, including two optimization stages, is presented in Algorithm \ref{alg3}. In the pre-optimization step of the NR task, a population $P_{NR}^{(0)}$ for the NR task is initialized randomly, and the general population-based NR algorithm $OptimizerNR$ can be employed to pre-optimize the NR task with a fixed number of function evaluations ($1-\lambda \times TFE_1$) (lines 1-2 in Algorithm \ref{alg3}). Then an initial network structure $X^{(0)}$ is selected from $P_{NR}^{(0)}$ by Algorithm \ref{alg1}, and the NR task on $X^{(0)}$ is pre-optimized by the general population-based CD algorithm $OptimizerCD_{pre}$ with a fixed number of function evaluations ($1-\lambda \times TFE_2$) (line 3-4 in Algorithm \ref{alg3}). After the pre-optimization stage, the initial network structure $X^{(0)}$ and community partition $C^{(0)}$ can be obtained. Next, a normal optimization stage is performed to get more accurate solutions by utilizing the common knowledge across the NR and CD tasks. The same algorithm as $OptimizerNR$ is employed in the main loop to optimize the NR task with a fixed number of function evaluations $N_1$ in the \textit{t} generation of NC (line 8 in Algorithm \ref{alg3}). Then a general population-based CD algorithm of dynamic network $OptimizerCD_{normal}$ is performed to optimize \ref{eq9} with a fixed number of function evaluations $t_2$ by utilizing the knowledge (network structure) acquired from the NR task for two consecutive generations (lines 10-11 in Algorithm \ref{alg3}). Next, knowledge transfer from the CD task to the NR task is employed to improve the convergence and diversity of the population $P_{NR}^{(t)}$ (line 12 in Algorithm \ref{alg3}). Finally, the decision-maker has selected the network structure $X^*$ and community partition $C^*$ (lines 17-18 in Algorithm \ref{alg3}).

\section{Experiments}
This section verifies the effectiveness of NC on the proposed test Suite. Firstly, we give the experimental settings in Section V.A. Then, the performance of our proposal is tested in Section V.B. Next, we analyze the effect of knowledge transfer of NC in Section V.C. Finally, in Section V.D, the parameter analysis is also given. We perform all experiments on a PC with Windows and Intel Core i7-8700 CPU at 3.20 GHz and 16GB RAM.

\subsection{Experimental Setup}

\begin{table} [t]
%\tiny
\centering
\caption{Test suite of MNRCDPS.}\label{tab1}
\resizebox{.5\textwidth}{!}{ 
\begin{tabular}{ccccccc}
\hline
ProblemID &  Network & $N_s-L$ & Type & $D$ & $N$ & $S_c$\\
\hline
EG1 &  ZK & \multirow{4}{*}{5-10} & \multirow{4}{*}{\makecell{Binary \\ $x_{ij} \in \{0,1\}$}} & 1156 & 34  & 2\\
EG2 &  polbooks &  &  & 11025 & 105 & 3\\
EG3 &  football &  &  & 13255 & 115 & 12\\
EG4 &  dolphin &  &  & 3844 & 62 & 2\\
\hline
EG5 &  ZK & \multirow{4}{*}{20-10} & \multirow{4}{*}{\makecell{Binary \\ $x_{ij} \in \{0,1\}$}} & 1156 & 34 & 2\\
EG6 &  polbooks &  &  & 11025 & 105 & 3\\
EG7 &  football &  &  & 13255 & 115 & 12\\
EG8 &  dolphin &  &  & 3844 & 62 & 2\\
\hline
EG9 &  BA & \multirow{4}{*}{20-10} & \multirow{4}{*}{\makecell{Binary \\ $x_{ij} \in \{0,1\}$}} & 2500 & 50 & -\\
EG10 &  ER &  &  & 2500 & 50 & -\\
EG11 &  NW &  &  & 2500 & 50 & -\\
EG12 &  WS &  &  & 2500 & 50 & -\\
\hline
RN1 &  ZK & \multirow{4}{*}{5-10} & \multirow{4}{*}{\makecell{Binary \\ $x_{ij} \in \{0,1\}$}} & 1156 & 34 & 2\\
RN2 &  polbooks &  &  & 11025 & 105 & 3\\
RN3 &  football &  &  & 13255 & 115 & 12\\
RN4 &  dolphin &  &  & 3844 & 62 & 2\\
\hline
RN5 &  ZK & \multirow{4}{*}{20-10} & \multirow{4}{*}{\makecell{Binary \\ $x_{ij} \in \{0,1\}$}} & 1156 & 34 & 2\\
RN6 &  polbooks &  &  & 11025 & 105 & 3\\
RN7 &  football &  &  & 13255 & 115 & 12\\
RN8 &  dolphin &  &  & 3844 & 62 & 2\\
\hline
RN9 &   BA & \multirow{4}{*}{20-10} & \multirow{4}{*}{\makecell{Binary \\ $x_{ij} \in \{0,1\}$}} & 2500 & 50 & -\\
RN10 &  ER &  &  & 2500 & 50 & -\\
RN11 &  NW &  &  & 2500 & 50 & -\\
RN12 &  WS &  &  & 2500 & 50 & -\\
\hline
\end{tabular}}
\end{table}

\paragraph{Test Suite}
We employ the commonly used networks to construct the test suite, and these networks represent most types of networks. In this section, four real social networks, football \cite{newman2006finding}, polbooks, dolphin \cite{newman2006finding}, and ZK \cite{zachary1977information} are employed to construct the test suite of MNRCDPs. Meanwhile, synthetic networks, such as Erd{\H{o}}s–R{\'e}nyi random networks (ER)  \cite{erdHos1960evolution}, Barab{\'a}si–Albert scale-free networks (BA) \cite{barabasi1999emergence},
Newman–Watts small-world networks (NW) \cite{newman1999renormalization}, and Watts–Strogatz small-world networks (WS) \cite{watts1998collective} with the average degree 6, are employed to enrich the property of the designed benchmark. These synthetic networks may not have a good network division. The details of those networks are presented in Table \ref{tab1}, including the type of variables for the NR task, the number of variables $D$ for the NR task, the number of nodes $N$, the number of links $Li$, and the number of communities $S_c$. In terms of $D$, these problems are high-dimensional. In the test suite, two types of response sequences are generated from the given networks by the methods described in the Appendix. The first one is the case of 5 response sequences with ten rounds each ($N_S$=5, $L$=10). The second is the case of 20 response sequences with ten rounds each ($N_S$=20, $L$=10). In Table \ref{tab1}, EG1–EG12 and RN1–RN12 represent the EG and RN tasks, respectively. This test suite is widely used in \cite{han2015robust,wang2011network}.

\begin{table*} [t]
%\tiny
\centering
\caption{Experimental results
(average(standard deviation)) of NC-NSGA-II, NC-SPEA2, NR2CD-NSGA-II, NR2CD-SPEA2, NC-SparseEA, and NR2CD-SparseEA on EG1-EG8 and RN1-RN8 in terms of MCC (NR task) and NMI (CD task). $-/\approx/+$ represents loss/tie/win.}
\label{tab3}
\resizebox{\textwidth}{!}{
\begin{tabular}{cc|cc|cc|cc}
\hline
ID & Type of Tasks & NR2CD-NSGA-II & NC-NSGA-II & NR2CD-SPEA2 & NC-SPEA2 & NR2CD-SparseEA & NC-SparseEA\\
\hline
\multirow{2}{*}{EG1} &  NR Task & 7.87e-01(6.60e-03)$-$ & \textbf{8.39e-01(1.73e-02)} & 7.98e-01(2.31e-02)$-$ & \textbf{8.37e-01(2.38e-02)} & 7.88e-01(3.95e-03)$-$ & \textbf{8.56e-01(3.56e-02)} \\
 & CD Task & 4.25e-01(8.93e-02)$-$ & \textbf{9.18e-01(1.16e-01)} & 4.76e-01(1.52e-01)$-$ & \textbf{9.75e-01(4.38e-02)} & 4.16e-01(8.65e-02)$-$ & \textbf{6.65e-01(1.91e-01)} \\
\hline
\multirow{2}{*}{EG2} &  NR Task & 5.22e-01(1.87e-03)$-$ & \textbf{5.41e-01(6.27e-03)} & 5.19e-01(8.61e-03)$\approx$ & \textbf{5.19e-01(6.80e-03)} & 5.80e-01(6.19e-03)$-$ & \textbf{6.48e-01(6.27e-03)}\\
 & CD Task & 3.03e-01(3.93e-03)$-$ & \textbf{3.52e-01(5.22e-03)}	& 3.10e-01(1.85e-02)$-$ & \textbf{3.27e-01(7.07e-03)} & 8.75e-02(1.51e-02)$-$ & \textbf{3.52e-01(3.45e-04)} \\
\hline
\multirow{2}{*}{EG3} &  NR Task & 5.14e-01(6.00e-03)$-$ & \textbf{5.24e-01(2.61e-03)} & 5.05e-01(5.92e-03)$-$ & \textbf{5.09e-01(2.25e-03)} & 5.60e-01(2.36e-03)$-$ & \textbf{5.97e-01(5.08e-03)} \\
 & CD Task & 2.18e-01(3.78e-01)$-$ & \textbf{6.03e-01(6.48e-03)} & 1.74e-01(3.01e-01)$-$ & \textbf{5.27e-01(3.36e-02)} & 4.01e-01(3.55e-01)$-$ & \textbf{6.53e-01(2.37e-03)} \\
\hline
\multirow{2}{*}{EG4} &  NR Task & 6.77e-01(8.86e-03)$-$ & \textbf{7.32e-01(2.56e-02)} & 6.10e-01(5.00e-03)$-$ & \textbf{6.78e-01(1.56e-02)} & 6.45e-01(1.21e-02)$-$ & \textbf{7.47e-01(2.21e-02)} \\
 & CD Task & 2.77e-01(1.27e-01)$-$ & \textbf{7.13e-01(1.74e-01)} & 3.23e-01(5.16e-02)$-$ & \textbf{5.39e-01(1.48e-01)} & 2.67e-01(4.13e-02)$-$ & \textbf{6.33e-01(5.53e-02)} \\
\hline
\multirow{2}{*}{EG5} &  NR Task & 8.55e-01(3.28e-02)$-$ & \textbf{9.40e-01(4.83e-03)} & 8.44e-01(1.71e-02)$-$ & \textbf{9.08e-01(2.46e-02)} & 8.71e-01(1.26e-02)$-$ & \textbf{9.46e-01(1.10e-02)} \\
 & CD Task & 7.57e-01(1.38e-01)$-$ & \textbf{8.57e-01(1.35e-01)} & 4.12e-01(8.07e-02)$-$ & \textbf{8.89e-01(1.58e-01)} & 3.28e-01(4.90e-02)$-$ & \textbf{5.92e-01(1.90e-01)} \\
\hline
\multirow{2}{*}{EG6} &  NR Task & 5.28e-01(8.18e-04)$-$ & \textbf{5.63e-01(1.06e-03) }& 5.20e-01(2.09e-03)$-$ & \textbf{5.23e-01(8.94e-03)} & 6.35e-01(5.66e-03)$-$ & \textbf{7.04e-01(3.59e-03)}  \\
 & CD Task & 3.09e-01(1.18e-02)$-$ & \textbf{3.41e-01(1.95e-03)} & \textbf{3.20e-01(1.68e-02)}$+$ & 3.09e-01(1.83e-02) & 4.01e-01(3.15e-02)$-$ & \textbf{5.62e-01(3.21e-02)} \\
\hline
\multirow{2}{*}{EG7} &  NR Task & 5.16e-01(5.70e-03)$-$ & \textbf{5.25e-01(1.72e-02)} & \textbf{5.10e-01(3.23e-03)}$\approx$ & \textbf{5.07e-01(6.52e-03)} & 5.90e-01(1.71e-03)$-$ &\textbf{ 6.46e-01(2.32e-03)} \\
 & CD Task & 3.43e-01(2.97e-01)$-$ & \textbf{6.38e-01(1.38e-02)} & \textbf{5.50e-01(2.32e-02)}$+$ & 5.31e-01(3.93e-02) & 6.33e-01(5.29e-04)$-$ & \textbf{6.67e-01(1.44e-02)} \\
\hline
\multirow{2}{*}{EG8} &  NR Task & 7.19e-01(5.39e-03)$-$ & \textbf{8.21e-01(4.14e-03)} & 6.10e-01(8.18e-03)$-$ & \textbf{7.47e-01(4.01e-04)} & 7.38e-01(1.01e-03)$-$ & \textbf{8.45e-01(1.71e-02)} \\
 & CD Task & 4.80e-01(3.85e-02)$-$ & \textbf{7.00e-01(1.53e-01)} & 3.17e-01(7.62e-02)$-$ & \textbf{5.24e-01(1.71e-01)} & 3.28e-01(3.95e-02)$-$ & \textbf{7.74e-01(8.80e-02)} \\
\hline
\multirow{2}{*}{RN1} &  NR Task & 8.73e-01(9.07e-03)$-$ & \textbf{9.31e-01(1.95e-02)} & 8.58e-01(1.18e-02)$-$ & \textbf{8.92e-01(9.71e-03)} & 8.36e-01(8.07e-03)$-$ & \textbf{9.05e-01(7.51e-03)} \\
 & CD Task & 2.09e-01(3.82e-02)$-$ & \textbf{5.90e-01(1.86e-01)} & 3.22e-01(1.84e-01)$-$ & \textbf{5.77e-01(3.21e-03)} & 3.25e-01(5.08e-02)$-$ & \textbf{4.45e-01(1.22e-01)} \\
\hline
\multirow{2}{*}{RN2} &  NR Task & 5.05e-01(7.89e-03)$-$ & \textbf{5.30e-01(5.80e-03)} & 5.15e-01(7.19e-03)$\approx$ & \textbf{5.18e-01(6.80e-03)} & 5.61e-01(3.92e-03)$-$ & \textbf{6.19e-01(1.04e-02)} \\
 & CD Task & 3.01e-01(1.71e-02)$-$ & \textbf{3.46e-01(2.17e-03)} & 3.08e-01(4.31e-03)$-$ & \textbf{3.41e-01(5.38e-03)} & 1.94e-01(1.79e-01)$-$ & \textbf{3.50e-01(2.77e-03)}  \\
\hline
\multirow{2}{*}{RN3} &  NR Task & 5.05e-01(1.06e-02)$-$ & \textbf{5.15e-01(5.18e-04)} & 5.04e-01(3.32e-03)$\approx$ & \textbf{5.06e-01(3.55e-03)} & 5.48e-01(4.73e-03)$-$ & \textbf{5.96e-01(1.17e-02)} \\
 & CD Task & 3.66e-01(3.17e-01)$-$ & \textbf{6.28e-01(2.36e-03)} & 1.83e-01(3.17e-01)$-$ & \textbf{5.12e-01(1.97e-02)} & 6.04e-01(8.91e-03)$-$ & \textbf{6.28e-01(3.41e-03)} \\
\hline
\multirow{2}{*}{RN4} &  NR Task & 6.51e-01(9.06e-03)$-$ & \textbf{7.30e-01(2.59e-02)} & 6.09e-01(8.40e-03)$-$ & \textbf{6.61e-01(2.06e-02)} & 6.57e-01(1.31e-02)$-$ & \textbf{7.81e-01(1.27e-02)} \\
 & CD Task & 1.03e-01(8.76e-02)$-$ & \textbf{2.79e-01(1.12e-01)} & 6.58e-02(2.18e-02)$-$ & \textbf{2.27e-01(4.63e-02)} & 1.12e-01(4.30e-02)$-$ & \textbf{3.33e-01(6.72e-02)} \\
\hline
\multirow{2}{*}{RN5} &  NR Task & 8.88e-01(1.69e-02)$-$ & \textbf{9.48e-01(2.06e-03)} & 8.93e-01(2.25e-03)$-$ & \textbf{9.31e-01(1.24e-02)} & 8.89e-01(9.98e-03)$-$ & \textbf{9.59e-01(1.06e-03)} \\
 & CD Task & 4.93e-01(2.58e-01)$-$ & \textbf{7.65e-01(1.29e-02)} & 4.83e-01(1.07e-01)$-$ & \textbf{6.16e-01(2.30e-01)} & 3.18e-01(7.33e-02)$-$ & \textbf{4.07e-01(4.30e-02)} \\
\hline
\multirow{2}{*}{RN6} &  NR Task & 5.18e-01(5.60e-03)$-$ & \textbf{5.32e-01(3.10e-03)} & \textbf{5.13e-01(4.79e-03)}$\approx$ & \textbf{5.12e-01(1.67e-02)} & 5.78e-01(4.64e-03)$-$ & \textbf{6.57e-01(1.13e-02)} \\
 & CD Task & 2.95e-01(5.33e-03)$-$ & \textbf{3.39e-01(1.21e-02)} & 2.84e-01(6.15e-03)$-$ & \textbf{3.30e-01(2.47e-02)} & 8.02e-02(1.58e-02)$-$ & \textbf{3.48e-01(8.94e-04)} \\
\hline
\multirow{2}{*}{RN7} &  NR Task & 5.03e-01(2.75e-03)$\approx$ & \textbf{5.06e-01(4.31e-03)} & 5.04e-01(6.16e-03)$-$ & \textbf{5.12e-01(4.13e-03)} & 5.69e-01(3.18e-03)$-$ & \textbf{6.29e-01(2.60e-03)} \\
 & CD Task & 1.78e-01(3.08e-01)$-$ & \textbf{6.56e-01(7.54e-04)} & 1.86e-01(3.22e-01)$-$ & \textbf{5.51e-01(3.13e-02)} & 3.39e-01(3.38e-01)$-$ & \textbf{6.69e-01(2.10e-02)} \\
\hline
\multirow{2}{*}{RN8} &  NR Task & 6.94e-01(4.99e-03)$-$ & \textbf{7.69e-01(1.62e-02)} & 6.19e-01(6.56e-03)$-$ & \textbf{6.88e-01(1.78e-02)} & 7.02e-01(9.02e-04)$-$ & \textbf{8.53e-01(3.24e-03)} \\
 & CD Task & 5.96e-02(1.51e-02)$-$ & \textbf{3.40e-01(8.12e-02)} & 1.42e-01(8.09e-02)$-$ & \textbf{2.30e-01(4.72e-02)} & 1.80e-01(6.78e-02)$-$ & \textbf{4.81e-01(3.55e-02)} \\
\hline
$-/\approx/+$ &  NR Task & 15/1/0 &	$-$ & 11/5/0 & $-$ & 16/0/0 & $-$ \\
\hline
$-/\approx/+$ &  CD Task & 16/0/0 &	$-$ & 14/0/2 & $-$ & 16/0/0 & $-$ \\
\hline
\end{tabular}}
\end{table*}

\paragraph{Algorithms}
NSGA-II \cite{deb2002fast}, SPEA2 \cite{zitzler2001spea2}, and SparseEA \cite{tian2020an}, three state-of-the-art multiobjective EAs, are embedded in NC as OptimizerNR to form NC-NSGA-II,  NC-SPEA2, and NC-SparseEA, respectively. Then ECD \cite{liu2020detecting}, a state-of-the-art EA of the CD method in the dynamic network, and PNGACD \cite{liu2019evolutionary}, a state-of-the-art EA of the CD method that has the same base search operator as ECD, are considered as $OptimizerCD_{normal}$ and $OptimizerCD_{pre}$ in NC-NSGA-II,  NC-SPEA2, and NC-SparseEA respectively. All experiments are implemented on the multiobjective optimization platform PlatEMO \cite{tian2017platemo}.

In general, we should compare NC with the alternative of solving the tasks with a parallel method but not sharing any knowledge. To verify the performance of our proposal in the experiments, a comparison algorithm called NR2CD is designed, which follows the general process of the NR task and the CD task.

1) \textbf{Single-task NR Methods}. In NR2CD, we first reconstruct the network structure from the observed data and then detect the community partition where knowledge transfer does not occur across these two tasks. NSGA-II and SPEA2 are embedded in NR2CD as the base optimizer for the NRP to form NR2CD-NSGA-II and NR2CD-SPEA2, respectively. Because the first stage of NR2CD is the same as the single-task NR method, the results of the NR task in NR2CD can be used to finish this comparison. It should be highlighted that this paper aims to examine the potential for mutual promotion between NR and CD tasks, not to attain competitive network reconstruction performance. Therefore, we do not compare with those state-of-the-art approaches.

2) \textbf{Single-task CD Methods}. As shown in Fig. \ref{fig1}, none of the current methods can detect communities from dynamics directly. Thus, we cannot compare our proposal with the single task of the CD from dynamics. 

Different multitasking environments are created with synergistic and non-synergistic tasks to see the efficiency of solving simultaneous tasks. We consider the following two cases.

3) \textbf{CD Methods Aided by the NR Task}.
For a fair comparison, PNGACD is considered the base optimizer for the CD task in NR2CD-NSGA-II and NR2CD-SPEA2 and detects communities from the networks obtained by single-task NR methods. The results of the CD task in NR2CD can be used to finish this comparison.

4) \textbf{NR methods Aided by Communities}. CEMO-NR \cite{wu2020evolutionary} employed community partition to aid the task of network reconstruction. In NC, the idea of transferring the best communities for the NR task is inspired by CEMO-NR. The performance of NC matches that of CEMO-NR, and then we do not show the NR results of CEMO-NR.

5) \textbf{Evolutionary Multitasking Methods}. None of the current evolutionary multitasking methods can handle multitasking network reconstruction and community detection, especially their knowledge transfer strategies. The proposed NC is a unique design for this problem. Thus, we cannot compare our proposal with the state-of-the-art evolutionary multitasking methods.

\paragraph{Evaluation Metrics}
Two measurement indices are employed to evaluate the performance of our proposal, the Matthews correlation coefficient (MCC) and NMI. MCC is employed to measure the accuracy of the reconstructed network structure, and the NMI is employed to measure the quality of community partitions. 
For each case, all comparison algorithms are run 20 times independently to obtain statistical results, and the Wilcoxon rank-sum test is employed to test the significance of the experimental results.

\paragraph{Parameter Settings}
NSGA-II, SPEA2, and SparseEA used the same single-point crossover and bitwise mutation operators for the NR task. For the CD task, ECD and PNGACD use the same crossover, mutation, and migration operators in \cite{liu2020detecting}. The population size for the NR task and CD task is set to 100, and both the maximum number of function evaluations for the NR and CD tasks are assigned to 200000 in all experiments. In NC, $\lambda$ and $t_1$ are set to 0.5 and 1000, respectively. Then the number of time steps for the CD task in NC, marked as T, is $\left \lceil \lambda \times TFE_1/(N_1+2\times t_1) \right \rceil$ and the number of function evaluations for the CD task in each time step in NC, marked as $t_2$, is $\lambda TFE_2/T$. The Sub-population size $\alpha$ for the knowledge transfer from the CD task to the NR task in NC is set to 20. 
The details of parameter settings are listed in Table \ref{tab2}.

\begin{table} [t]
%\tiny
\scriptsize
\centering
\caption{Parameter settings.}\label{tab2}
\begin{tabular}{lll}
\hline
Parameters &  Value & Descriptions \\
\hline
$N_1$ & 100 & The population size for the NR task\\
$N_2$ & 100 & The population size  for the CD task \\
$TFE_1$ & 200000 & \makecell[l]{The maximum number of function evaluations for the \\ NR task} \\
$TFE_2$ & 200000 & \makecell[l]{The maximum number of function evaluations for the \\ CD task} \\
$\lambda$ & 0.5 & \makecell[l]{The share of the TFE used for the normal optimization \\ stage in NC} \\
$t_1$ & 1000 & \makecell[l]{The number of function evaluations for the knowledge \\ transfer from  the CD task to NR task in NC} \\
$P_c$ & 1 & \makecell[l]{The probabilities of crossover in NSGAII and SPEA2} \\
$P_m$ & $1/D$ & \makecell[l]{The probabilities of mutation in NSGAII and SPEA2} \\
$P_{mu}$ & 0.2 & \makecell[l]{The probabilities of mutation in ECD and PNGACD} \\
$P_{mi}$ & 0.2 & \makecell[l]{The probabilities of migration in ECD and PNGACD} \\
$P_{mu/mi}$ & 0.5 & \makecell[l]{The parameter to control the execution of mutation \\ and migration in ECD and PNGACD} \\
\hline
\end{tabular}
\end{table}

\subsection{Results and Discussion}
Table \ref{tab3} lists the mean and standard deviation of MCC (NR Task) and NMI (CD Task) obtained by different methods on problems EG1-EG8 and RN1-RN8 over 20 independent runs, where symbols “-”, “$\approx$” and “+” imply that the NR2CD-Alg is significantly worse, similar and better than NC-Alg on the Wilcoxon rank-sum test with 95\% confidence level, respectively. We mark the best results in boldface.

In Table \ref{tab3}, NC-Alg has shown excellent performance in terms of MCC and NMI in the test suite. Specifically, compared to the single NR task, NC-Alg performs better on 42, ties 6, and loses once out of 48 tasks in terms of MCC (NR Task), which demonstrates that the transferred community information in our method may be helpful. NC-Alg exceeds the NR2CD-Alg in 46 and loses 2 out of 48 cases in terms of NMI, which shows our proposal is better than the CD methods aided by the NR task. In NC-Alg, useful knowledge is transferred across the NR and CD tasks to improve each task’s performance. Since NC-Alg has the same basic evolutionary solver of the NR task and CD task as NR2CD-Alg, knowledge transfer’s effectiveness in NC-Alg is confirmed. For the problems with the polbooks network, NC-Alg and NR2CD-Alg have similar performances in terms of the NR task. Since the community partition in the reconstructed imprecise network may have a large gap from the community partition in the existing network, knowledge transfer may fail in large-scale networks. Moreover, it is found that the number of communities is much larger than the real number of communities. The size of transferred communities is too small to offset its consumption of computational resources. 

Table \ref{tab3} shows that the quality of both networks and communities, except for ZK, is not impressive. NMI is often below 0.5, which means that the approach cannot recover the true communities, but the MCC is not high. It can be found that the accuracy of the obtained network structure is not high, and the community detected on this network may deviate from the real community partition. This phenomenon appears since we do not design the specified operator for these two tasks, which may improve the performance violently. These experiments aim to verify the effectiveness of the designed knowledge transfer operator and whether NC can handle the task of CD and NR from dynamics jointly. We think the experiments have achieved these goals.

In the worst case, one network contains a community. In this case, NC can be considered as the single task of reconstructing the network alone. Moreover, it is challenging to create non-synergistic tasks since the more accurate the given network is, the more accurate the detection results are. Also, due to the limited ability of evolutionary algorithms, the performance of the NR decreases with bigger networks, as shown in Table \ref{tab3}. The core of the experiments is to verify whether the proposed knowledge transfer strategy is effective and whether NC can handle multitasking network reconstruction and community detection.

\begin{figure}
\centering
\includegraphics[width=0.4\textwidth]{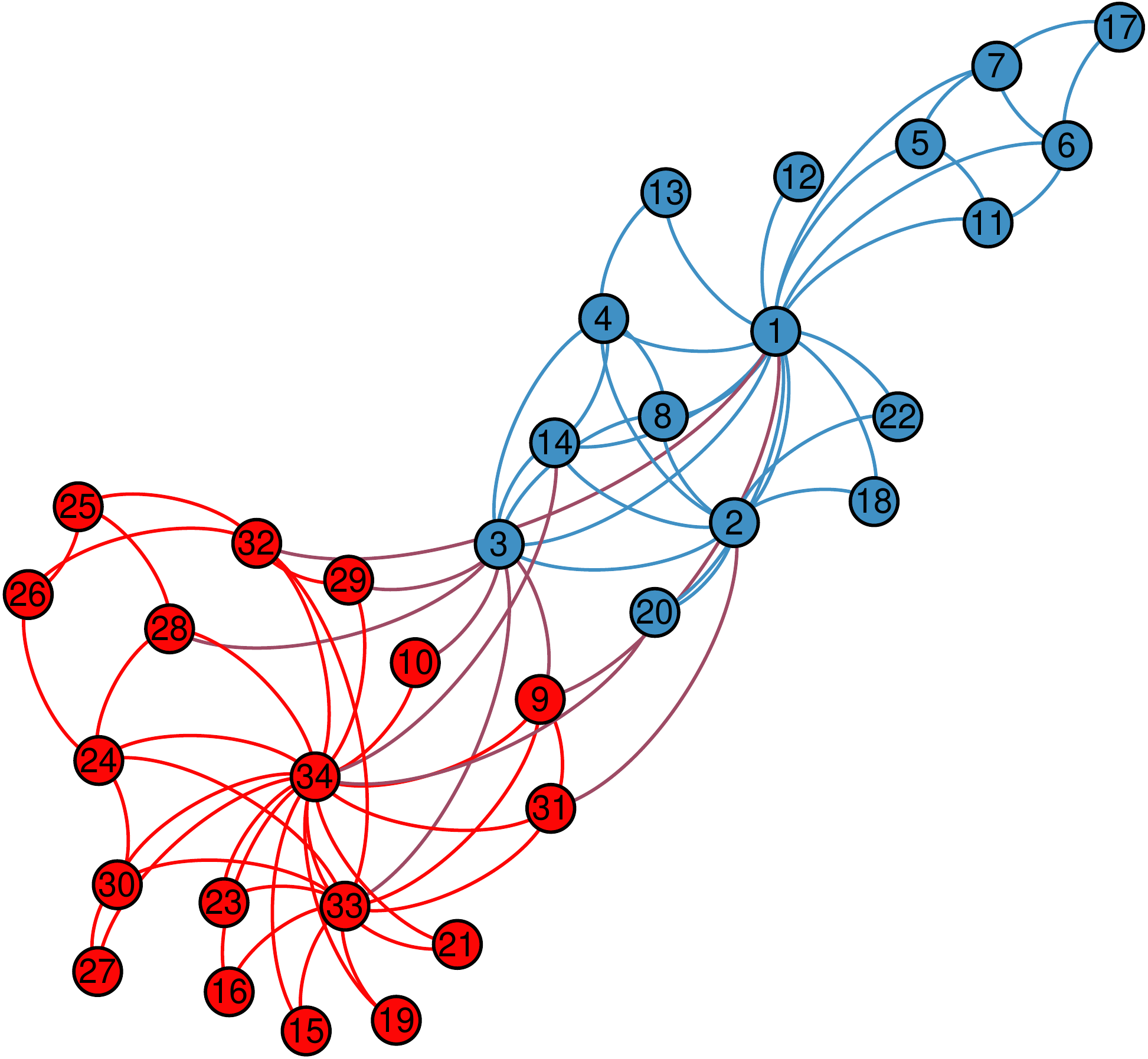}
\caption{An illustrative example of the community partition of the ZK network.}\label{fig5}
\end{figure}

To better illustrate the performance of NC-Alg, we visualize the network structure and community partition obtained by NC-NSGA-II on the ZK network. In Fig. \ref{fig5}, we mark the different communities with different colors. According to the connection between every two nodes obtained by knowledge transfer from the NR task to the CD task, NC-NSGA-II divides all nodes into two communities, which is consistent with the actual community division as shown in Table \ref{tab1}. Our proposal organizes the network into two divisions according to the connection density between nodes, where closely connected nodes are divided into the same community.

\begin{figure}[t]
	\centering
	\subfloat[] {\includegraphics[width=0.25\textwidth]{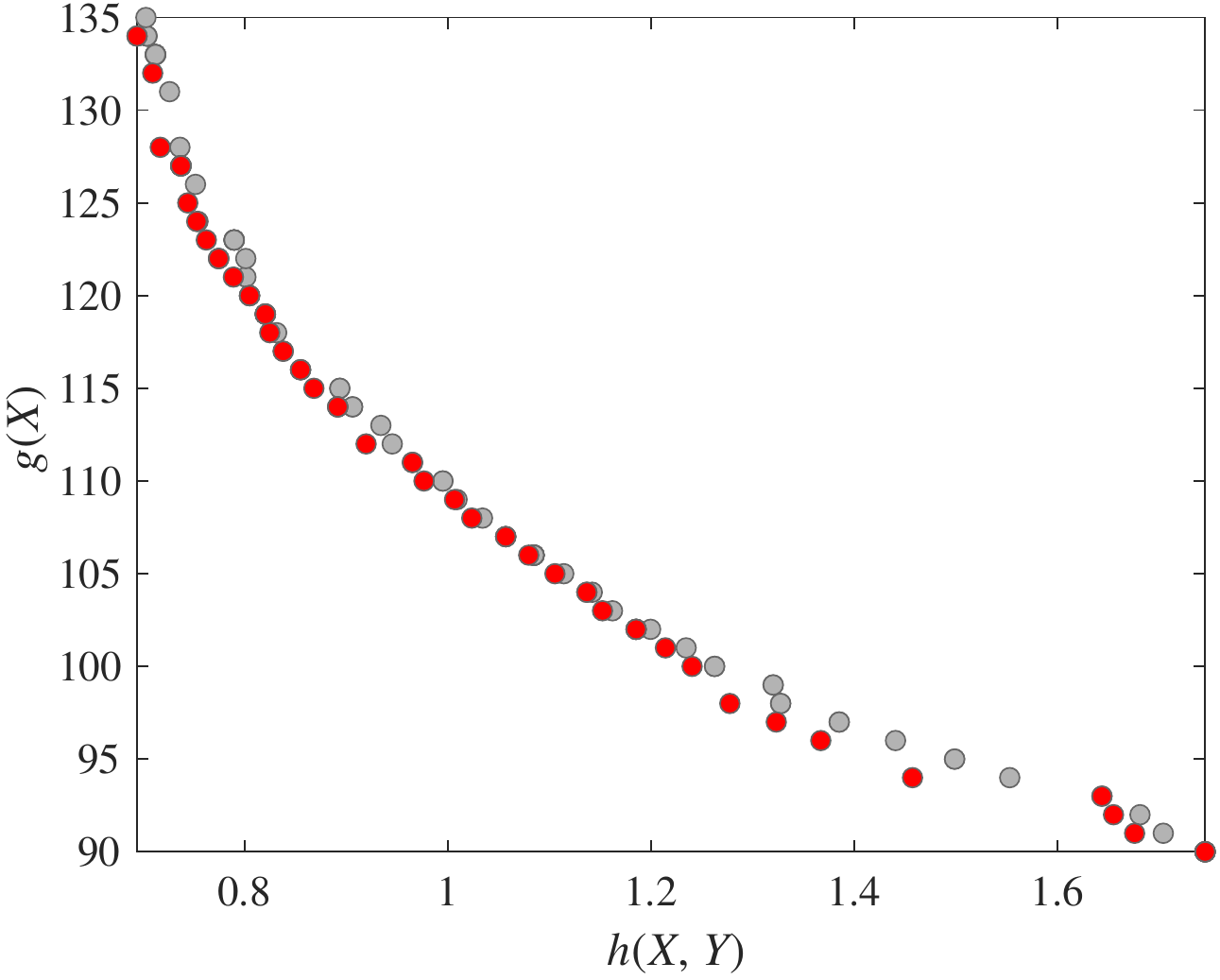}\label{fig6(a)}}
	\subfloat[] {\includegraphics[width=0.25\textwidth]{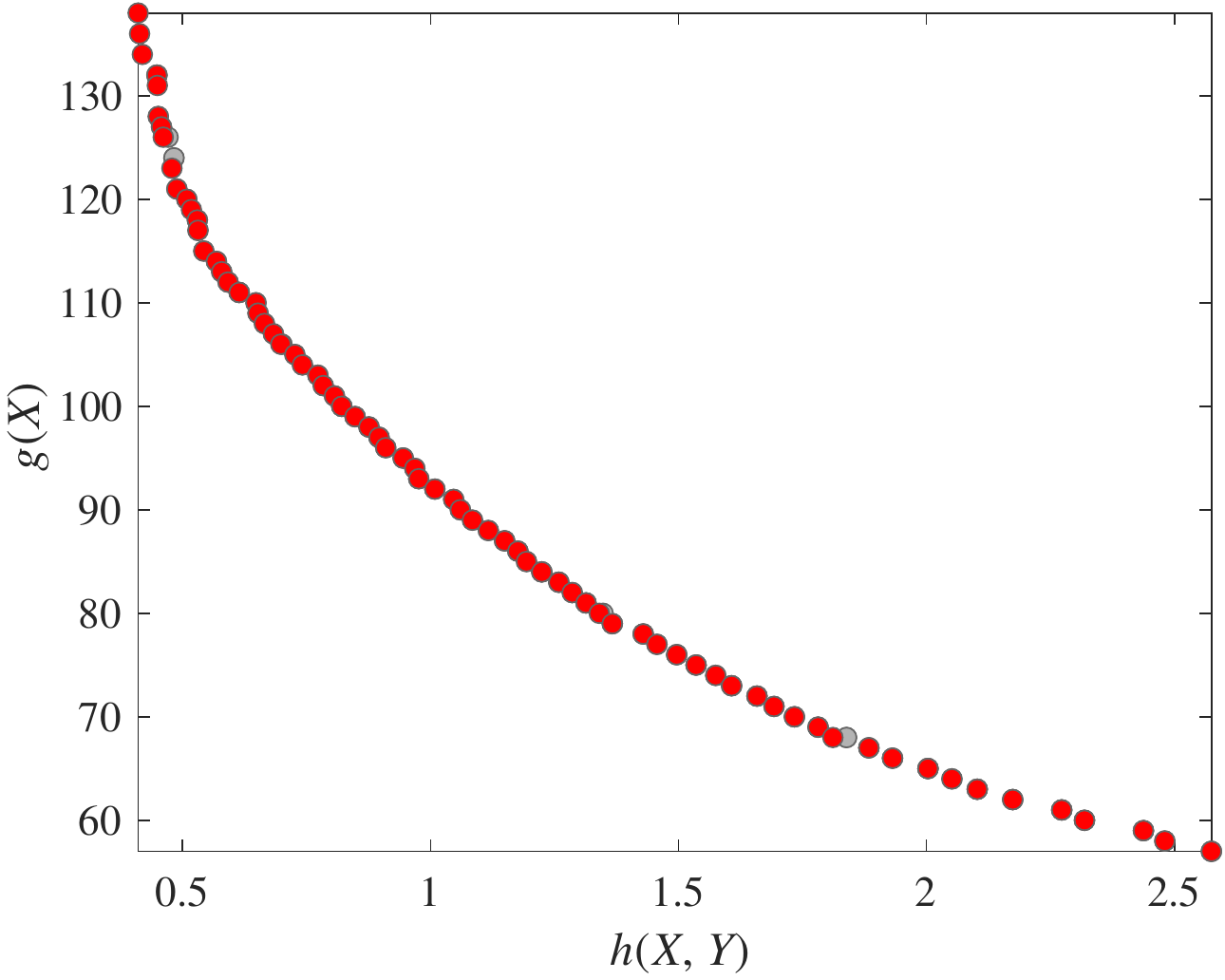}\label{fig6(b)}}
	\caption{The obtained non-dominated solution set before (grey) and after (red) the knowledge transfer from the CD task to the NR task. (a) The number of function evaluations for the NRP is 102100. (b) The number of function evaluations for the NRP is 198700.}
	\label{fig6}
\end{figure}

\subsection{Verification of Knowledge Transfer}
This section further illustrates the effectiveness of the knowledge transfer operators from the CD task to the NR task and from the NR task to the CD task in NC and takes the algorithm NC-NSGA-II and the problem EG1 as an example. 

\textbf{Knowledge transfer from the CD task to the NR task is valuable}. Fig. \ref{fig6}  shows the obtained non-dominated solution set before and after the knowledge transfer from the CD task to the NR task when the number of function evaluations for the NRP is 102100 and 198700. As can be observed in Fig. \ref{fig6(a)}, in the mid-stage of evolution, compared with the non-dominated solution set (grey) before knowledge transfer from the CD task to the NR task, the new non-dominated solution set (red) has better overall performance in terms of convergence and diversity, which illustrates the effectiveness of knowledge transfer from the CD task to NR task. As shown in Fig. \ref{fig6(b)}, in the later stages of evolution, the overall performance of the non-dominated solution set before and after the knowledge transfer is not much different in terms of convergence and diversity. Since the network structure obtained by the NR task has fully utilized the community partition obtained by the CD task, the knowledge transfer process from the CD task to the NR task in the later stage of evolution is not apparent for improving the non-dominated solution set.

\begin{figure}[t]
\centering
\includegraphics[width=0.35\textwidth]{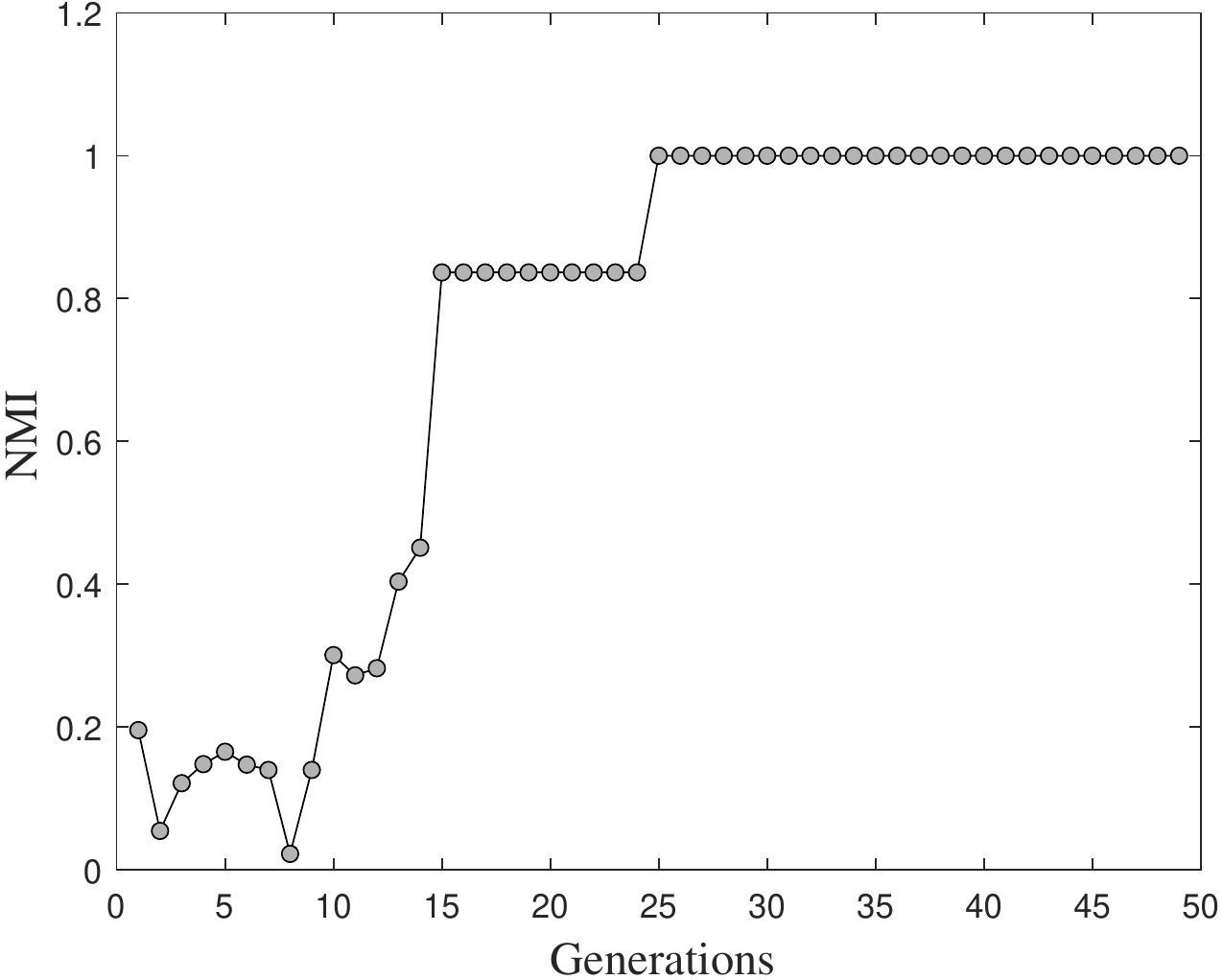}
\caption{A curve of the NMI values over generations, in which the NMI values are calculated based on community partitions at each generation and real community partitions.} \label{fig7}
\end{figure}

\textbf{Knowledge transfer from the NR task to the CD task is valuable}. Besides, Fig. \ref{fig7} shows a curve of values of NMI over time steps. NMI values are calculated by \ref{eq5} based on community partitions at each generation and the real community partition. As shown in Fig. \ref{fig7}, NMI shows an increasing trend over time, illustrating that knowledge from the NR task to the CD task positively impacts the CD task. We also find that negative knowledge may affect the performance shown in the early stage of evolution, but development can eliminate this phenomenon. The transfer process has little impact on the performance of our proposal at the last stage, but the knowledge transfer still uses the computational budget. We can decrease this negative knowledge transfer by reducing the frequency of knowledge transfer.

\subsection{Testing on Synthetic Networks}

We test the performance of the proposed algorithm under different characteristic networks, and the datasets used are shown in Table \ref{tab1} as EG9-EG12 and RN9-EG12. Since these problems do not have precise community partition, this section uses $Q$ as the performance metric for the CD task. See Table \ref{tab2} for other settings. In Table \ref{tab:syn}, it can be found that on eight problems, NC-NSGA-II performs better than NR2CD-NSGA-II. It can be shown that when the community partition is not apparent, our scheme is still effective and can also be adapted to a wide range of scenarios.%\ref{tab:EB}

\begin{table}[t]
 \centering
  \caption{Experimental results of NC-NSGA-II and NR2CD-NSGA-II on EG19-EG12 and RN9-RN12 in terms of MCC (NR task) and $Q$ (CD task).}
  \resizebox{.5\textwidth}{!}{ 
    \begin{tabular}{cccc}
    \toprule
    ProblemID & Task (Metric) & NR2CD-NSGA-II  & NC-NSGA-II \\
    \midrule
    \multirow{2}[2]{*}{EG9} & NR(MCC) & 6.22e-01(3.05e-01) & \textbf{7.92e-01(2.36e-02)} \\
          & CD($Q$) & 3.18e-01(1.71e-01) & \textbf{3.89e-01(0.00e+00)} \\
    \midrule
    \multirow{2}[2]{*}{EG10} & NR(MCC) & 6.96e-01(3.41e-01) & \textbf{9.49e-01(2.60e-02)} \\
          & CD($Q$) & 5.75e-01(2.88e-01) & \textbf{7.09e-01(3.82e-02)} \\
    \midrule
    \multirow{2}[2]{*}{EG11} & NR(MCC) & 8.47e-01(1.36e-02) & \textbf{9.73e-01(1.48e-02)} \\
          & CD($Q$) & 7.16e-01(4.38e-02) & \textbf{7.60e-01(5.07e-02)} \\
    \midrule
    \multirow{2}[2]{*}{EG12} & NR(MCC) & 7.11e-01(3.48e-01) & \textbf{9.58e-01(1.57e-02)} \\
          & CD($Q$) & 6.86e-01(3.38e-01) & \textbf{7.95e-01(2.57e-02)} \\
    \midrule
    \multirow{2}[2]{*}{RN9} & NR(MCC) & 6.70e-01(3.28e-01) & \textbf{9.35e-01(7.99e-03)} \\
          & CD($Q$) & 3.74e-01(1.89e-01) & \textbf{5.14e-01(5.94e-02)} \\
    \midrule
    \multirow{2}[2]{*}{RN10} & NR(MCC) & 7.27e-01(3.57e-01) & \textbf{9.29e-01(2.50e-02)} \\
          & CD($Q$) & 6.44e-01(3.17e-01) & \textbf{8.22e-01(5.44e-02)} \\
    \midrule
    \multirow{2}[2]{*}{RN11} & NR(MCC) & 8.66e-01(2.51e-02) & \textbf{9.45e-01(1.57e-02)} \\
          & CD($Q$) & 6.76e-01(9.35e-02) & \textbf{7.53e-01(3.39e-02)} \\
    \midrule
    \multirow{2}[2]{*}{RN12} & NR(MCC) & 7.23e-01(3.54e-01) & \textbf{9.43e-01(2.61e-02)} \\
          & CD($Q$) & 6.80e-01(3.45e-01) & \textbf{7.58e-01(3.42e-02)} \\
    \midrule
    \multirow{2}[2]{*}{$-/\approx/+$} & NR(MCC) & 8/0/0 & $-$ \\
          & CD($Q$) & 8/0/0 & $-$ \\
    \bottomrule
    \end{tabular}}
  \label{tab:syn}
\end{table}

% \begin{table}[htbp]
%  \centering
%   \caption{Experimental results of NC-NSGA-II and BNRCD on SIS1-SIS4 in terms of MCC (NR task) and $Q$ (CD task).}
%   \resizebox{.5\textwidth}{!}{ 
%     \begin{tabular}{cccc}
%     \toprule
%     ProblemID & Task (Metric) & NR2CD-NSGA-II  & BNRCD \\
%     \midrule
%     \multirow{2}[2]{*}{SIS1} & NR(MCC) & () & () \\
%           & CD($Q$) & () & () \\
%     \midrule
%     \multirow{2}[2]{*}{SIS2} & NR(MCC) & () & () \\
%           & CD($Q$) & () & () \\
%     \midrule
%     \multirow{2}[2]{*}{SIS3} & NR(MCC) & () & () \\
%           & CD($Q$) & () & () \\
%     \midrule
%     \multirow{2}[2]{*}{SIS4} & NR(MCC) & () & () \\
%           & CD($Q$) & () & () \\
%     \bottomrule
%     \end{tabular}}
%   \label{tab:EB}
% \end{table}

\subsection{Parameter Sensitivity}

The effects of two critical parameters are analyzed in this section, and we take the algorithm NC-NSGA-II and the problem EG1 as an example. Two key parameters are shown as follows: 1) $\lambda$, the share of the TFE used for the normal optimization stage in NC; 2) $t_1$, the number of function evaluations for the knowledge transfer from the CD task to the NR task. To analyze one parameter visually, values of other parameters are fixed.

\begin{figure}[t]
	\centering
	\subfloat[] {\includegraphics[width=.25\textwidth]{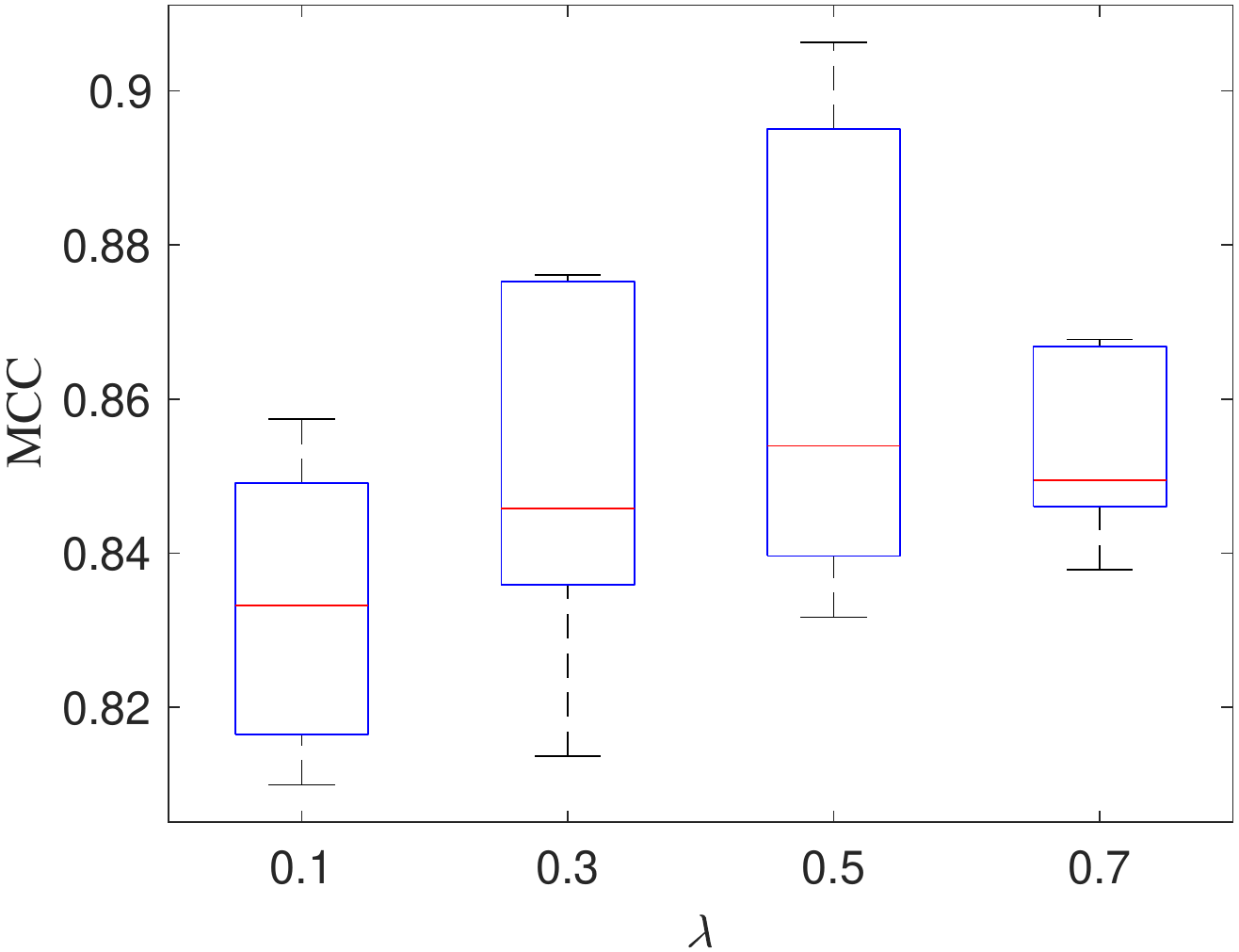}\label{fig8(a)}}
	\subfloat[] {\includegraphics[width=.25\textwidth]{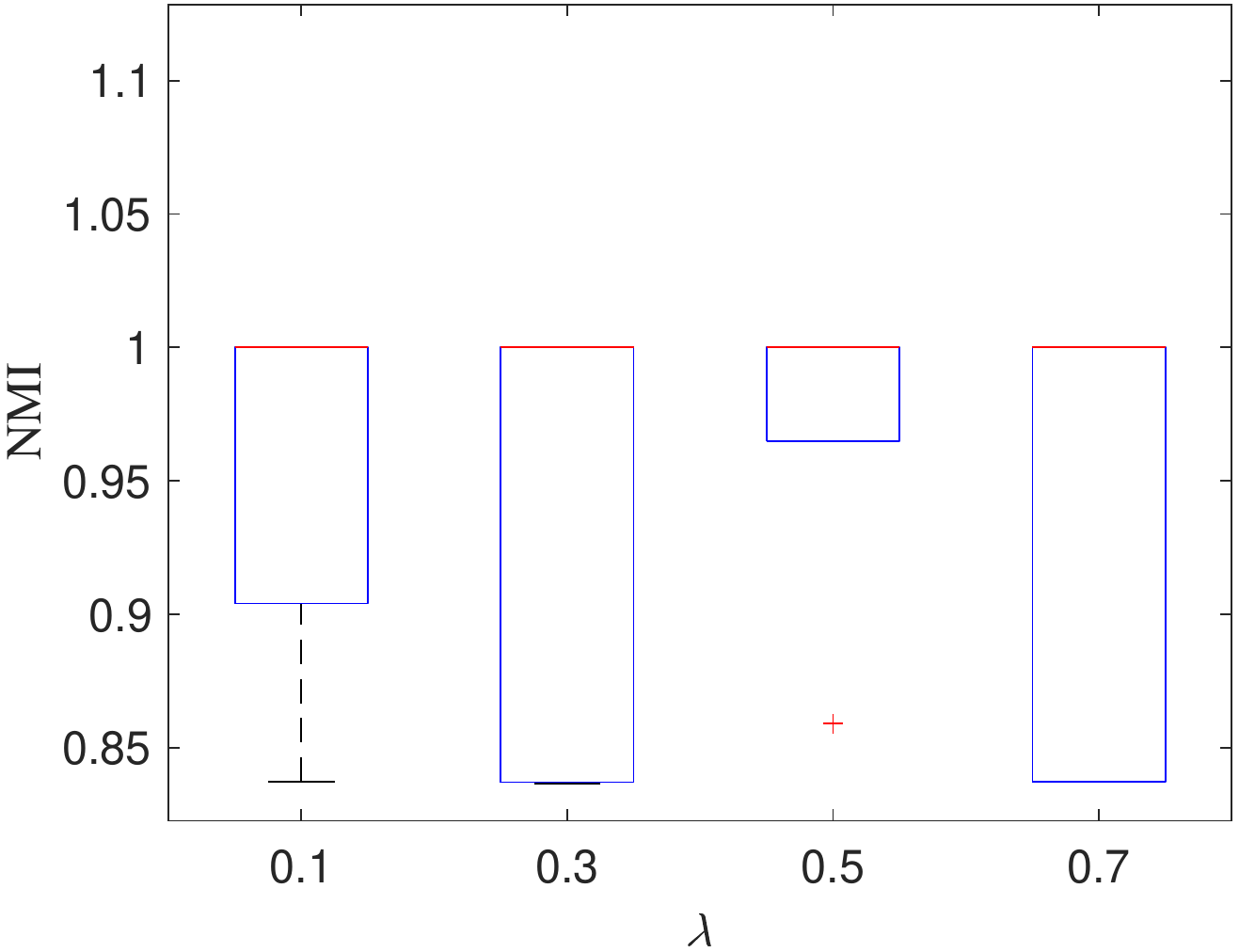}\label{fig8(b)}}
	\caption{The MCC and NMI versus varying $\lambda$. (a) MCC. (b) NMI.}
	\label{fig8}
\end{figure}

\textbf{Effect of $\lambda$ on NC}. Fig. \ref{fig8} shows the MCC and NMI versus varying $\lambda$. The value of $\lambda$ is set to 0.1, 0.3, 0.5, and 0.7. In Fig. \ref{fig8(a)}, the value of MCC increases with $\lambda$ until $\lambda>0.3$. Then with increasing $\lambda$, the median value of MCC decreases. Fig. \ref{fig8(b)} shows that when $\lambda$ is 0.5, the value of NMI is the most stable in all cases. With the increase of $\lambda$, the NMI and MCC first increase and then decrease. This case appears because the accuracy of initializing the network impacts knowledge transfer in the normal optimization stage. With the increase of $\lambda$, the $TFE$ used for the normal optimization stage decreases, which results in the inability to fully utilize the knowledge between the NR task and the CD task.

\begin{figure}[t]
	\centering
	\subfloat[] {\includegraphics[width=.25\textwidth]{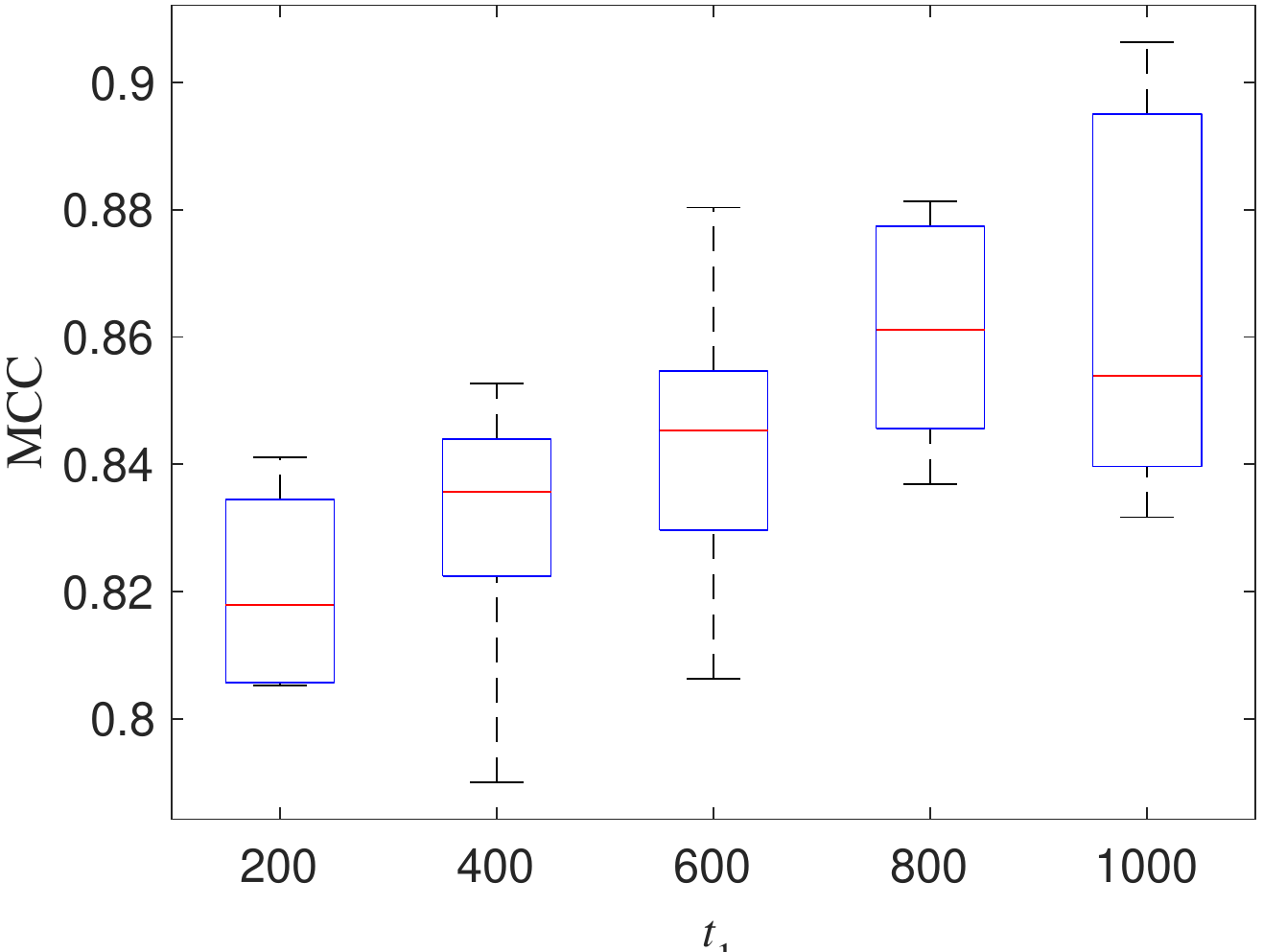}\label{fig9(a)}}
	\subfloat[] {\includegraphics[width=.25\textwidth]{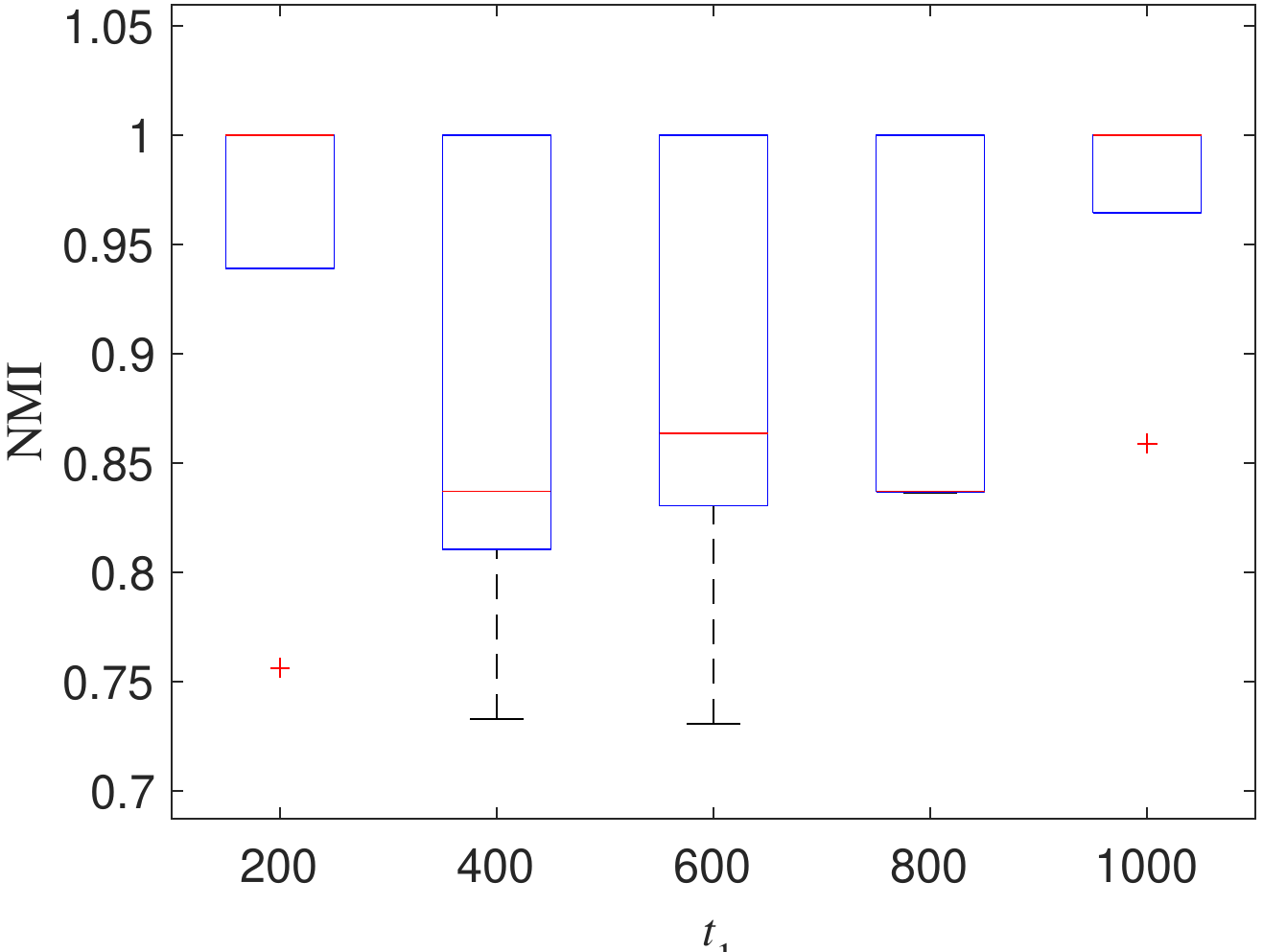}\label{fig9(b)}}
	\caption{The MCC and NMI versus varying $t_1$. (a) MCC. (b) NMI.}
	\label{fig9}
\end{figure}

\textbf{Effect of $t_1$ on NC}. Fig. \ref{fig9} shows the MCC and NMI versus varying $t_1$. The value of $t_1$ is set to 200, 400, 600, 800, and 1000. In Fig. \ref{fig9(a)}, with increasing $t_1$, NC obtains the greater value of MCC. In Fig. \ref{fig9(b)}, when $t_1$ is 1000, the median value of NMI is 1; that is, the community partition in the EG1 problem is completely identified. With increasing $t_1$, the transferred knowledge (network structure) can be more fully utilized to assist the NR task.

It can be seen from the observation that these parameters significantly impact the performance of NC. These parameters affect the frequency of knowledge transfer and the convergence and diversity of the obtained non-dominated solution. Since different NR and CD problems have unique properties, the collection of parameters in the problem EG1 may not suit the others.

\section{Conclusions}
This paper answers the question: could the joint optimization of NR and CD tasks effectively improve these two tasks’ performance? The answer is Yes, which is ensured by the proposed \emph{Network Collaborator}. The core is to determine what knowledge should be transferred across two tasks. Benefiting from the proposed evolutionary multitasking framework, we explicitly transfer the better community partition obtained by the CD task to aid the NR task and share the better network structure obtained by the NR task to aid the CD task. The experimental results on 24 cases show that this strategy is effective and supports the claim that joining these two tasks has a synergistic effect. The discovery of communities significantly improves the reconstruction accuracy, which finds a better community partition to perform these tasks in isolation. However, in terms of large-scale complex systems, the NR task’s low accuracy will decrease the CD task’s performance. In the future, more robust NR methods should be further studied. Besides, applying our proposal to find overlapping communities is a promising research topic in future work.

\section*{CRediT authorship contribution statement}
\textbf{Kai Wu}: Conceptualization, Supervision, Writing - original draft. \textbf{Chao Wang}: Methodology, Software, Validation, Writing - original draft. \textbf{Junyuan Chen}: Formal analysis, Writing - review \& editing. \textbf{Jing Liu}: Writing - review \& editing, Funding acquisition.

\section*{Declaration of competing interest}
The authors declare that they have no known competing financial interests or personal relationships that could have appeared to
influence the work reported in this paper.

\section*{Data availability}
Data will be made available on request.

\section*{Acknowledgments}
This work was supported in part by the National Natural Science Foundation of China under Grant 62206205, in part by the Guangdong High-level Innovation Research Institution Project under Grant 2021B0909050008, and in part by the Guangzhou Key Research and Development Program under Grant 202206030003.

\section*{Appendix}

\textbf{EG Problem}. The evolutionary game (EG) model has commonly been used to model node-to-node interactions in complex systems. In each round of the EG model, two players should choose a cooperation strategy or defection. Then the payoffs of these players are determined by the game's strategy and payoff matrix. The prisoner's dilemma games \cite{szabo2007evolutionary} are used in this paper. Its payoff matrix is denoted as follows:
\begin{equation}
P = \left( {\begin{array}{*{20}{c}}
1&0\\
{1.2}&0
\end{array}} \right)
\end{equation}

If both choose different strategies, the defector gets a reward of 1.2, and the cooperator receives a prize of 0. And if both decide to cooperate (or defect), all players get rewards 1 (or 0). Formally, for player \textit{i} in round \textit{t}, its payoff is as follows:

\begin{equation}\label{eq11}
    {Y_i}(t) = \sum\limits_{j = 1}^N {{x_{ij}}S_i^{\rm{T}}(t){P}{S_j}(t)} 
\end{equation}
where $Si(t)$ is the strategy of player \textit{i} in the \textit{t}-th round and \textit{T} represents “transpose”. $x_{ij}=1$ if players \textit{i} and \textit{j} are connected and $x_{ij}=0$ otherwise. To maximize their payoff at the next round, players adjust strategies according to their payoff and their neighbors after each round of the game. Fermi rule \cite{nowak1992evolutionary} is adopted to update the strategy in our simulations, which is as follows:
\begin{equation}\label{eq12}
    W\left( {{S_i} \leftarrow {S_j}} \right) = \frac{1}{{1 + \exp \left[ {\left( {{Y_i} - {Y_j}} \right)/\kappa } \right]}}
\end{equation}
where $\kappa=0.1$. To find the connections among players, the EG problem can be expressed as follows:
\begin{equation}
\begin{array}{l}
\mathop {\min }\limits_X F = \left( {h\left( {X,Y} \right) = \sum\limits_{i = 1}^N {\left\| {{U_i}{X_i} - {Y_i}} \right\|_2^2} ,g(X) = {{\left\| X \right\|}_0}} \right)\\
s.t.\;\left\{ {\begin{array}{*{20}{c}}
\begin{array}{l}
{X_i} = {\left[ {{x_{i1}},{x_{i2}}, \ldots ,{x_{iN}}} \right]^T} \in {\left\{ {0,1} \right\}^N}\\
{Y_i} = {[{Y_i}\left( 1 \right),{Y_i}\left( 2 \right), \ldots ,{Y_i}\left( L \right)]^T}
\end{array}\\
{{U_i} = \left( {\begin{array}{*{20}{c}}
{S_i^{\rm T}(1){P}{S_1}(1)}& \ldots &{S_i^{\rm T}(1){P}{S_N}(1)}\\
 \vdots & \ddots & \vdots \\
{S_i^{\rm T}(L){P}{S_1}(L)}& \cdots &{S_i^{\rm T}(L){P}{S_N}(L)}
\end{array}} \right)}
%\end{array}\;,\;i = 1,2, \ldots ,N}
\end{array}}
\right.
\end{array}
\end{equation}
where \textit{L} is the number of rounds. The first goal is to minimize the difference between the real payoff data and the generated payoff data of all players, and the second goal is to ensure the sparsity of the learned EG network. The simulation of the EG is described as follows \cite{wang2011network}:
\begin{enumerate}
    \item Input an EG network with nodes representing players;
    \item Each player chooses cooperation or defection;
    \item Calculate the payoff of player $i$ by (\ref{eq11});
    \item Update the strategy of player $i$ by (\ref{eq12});
    \item Repeat Step 3) to Step 4) $T$ times.
\end{enumerate}

For this dynamic process, the strategies and the payoffs of all players in different rounds are recorded as observational data. For this dynamic process, the strategies and the payoffs of all players in different rounds are recorded as observational data.

\textbf{RN Problem} \cite{han2015robust}. Resistor network dynamics is a standard circuit system considering current transportation in the resistor. The resistance of a resistor across nodes \textit{i} and \textit{j} is denoted as $r_{ij}$. For simplicity, $r_{ij}=1$ if \textit{i} and \textit{j} are directly connected by a resistor and $r_{ij}=\infty$ otherwise. According to Kirchhoff’s laws at different periods, assuming the voltages at the nodes and resistances of connections are known, the currents at the nodes are as follows:
\begin{equation}\label{eq14}
    {I_i}\left( t \right) = \sum\limits_{j = 1}^N {\frac{1}{{{r_{ij}}}}\left( {{V_i}\left( t \right) - {V_j}\left( t \right)} \right)}
\end{equation}
where $I_i$ is the total current at node \textit{i} and $V_i=V*sin[(w+\triangle w_i)t]$ is the voltage. In this paper, $V^*=1$ is the voltage peak, $w=10^3$ is the frequency, and $\triangle w_i\in[0, 20]$ is the perturbation. Assume that only the voltages and currents at the nodes are measurable, and the resistor network can be reconstructed as follows:
\begin{equation}
\begin{array}{l}
\mathop {\min }\limits_X F = \left( {h\left( {X,Y} \right) = \sum\limits_{i = 1}^N {\left\| {{R_i}{X_i} - {Y_i}} \right\|_2^2} ,g(X) = {{\left\| X \right\|}_0}} \right)\\
s.t.\;\left\{ {\begin{array}{*{20}{c}}
{{X_i} = {{\left[ {{x_{i1}}{\rm{ = }}\frac{{\rm{1}}}{{{r_{i1}}}},{x_{i2}}{\rm{ = }}\frac{{\rm{1}}}{{{r_{i2}}}}, \ldots ,{x_{iN}}{\rm{ = }}\frac{{\rm{1}}}{{{r_{iN}}}}} \right]}^T} \in {{\left\{ {0,1} \right\}}^N}}\\
{{Y_i} = {{[{I_i}\left( 1 \right),{I_i}\left( 2 \right), \ldots ,{I_i}\left( L \right)]}^T}}\\
{{R_i} = \left( {\begin{array}{*{20}{c}}
{{V_i}(1) - {V_1}(1)}& \ldots &{{V_i}(1) - {V_N}(1)}\\
 \vdots & \ddots & \vdots \\
{{V_i}(L) - {V_1}(L)}& \cdots &{{V_i}(L) - {V_N}(L)}
\end{array}} \right)}
%\end{array},\;i = 1,2, \ldots ,N}
\end{array}}
\right.
\end{array}
\end{equation}
where \textit{L} is the rounds of the observation data. The first goal is to minimize the difference between the real current data and the generated current data, and the second goal is to ensure the sparsity of the RN. The simulation of the RN is described as follows:
\begin{enumerate}
    \item Input an RN network;
    \item Each node state is obtained from a random number $\triangle w_i\in[0, 20]$;
    \item Calculate the voltage of node \textit{i} by $V_i=V^*sin[(w+\triangle w_i)t]$;
    \item Calculate the electrical current of the node \textit{i} by (\ref{eq14});
    \item Repeat Step 3) to Step 4) $T$ times.
\end{enumerate}

The voltages and the currents at the nodes simultaneously are recorded as observational data for this dynamic process.

% Uncomment and use as the case may be
%\begin{theorem} 
%\end{theorem}

% Uncomment and use as the case may be
%\begin{lemma} 
%\end{lemma}


\printcredits

\bibliographystyle{cas-model2-names}

\bibliography{cas-refs}



\end{document}